\documentclass{aa}  

\usepackage{graphicx}
\usepackage{txfonts}
\usepackage{subcaption}

\usepackage{gensymb}

\usepackage[thinc]{esdiff}

\usepackage[version=4]{mhchem}
\usepackage{chemformula}

\AddToHook{begindocument/before}{\usepackage{hyperref}}

\begin{document}

   \title{Zooming into the water snowline: high resolution water observations of the HL Tau disk}

   \subtitle{}

   \author{M. Leemker \inst{1}
          \and
          S. Facchini \inst{1}
          \and
          P. Curone \inst{2}
          \and
          L. Rampinelli \inst{1}
          \and 
          M. Benisty \inst{3}
          \and 
          A. Garufi \inst{4}
          \and 
          E. Humphreys \inst{5}
          }

        \institute{Dipartimento di Fisica, Università degli Studi di Milano, Via Celoria 16, 20133 Milano, Italy\\
              \email{margot.leemker@unimi.it}
         \and
         Departamento de Astronom\'ia, Universidad de Chile, Camino El Observatorio 1515, Las Condes, Santiago, Chile
         \and
         Max-Planck-Institut für Astronomie, Königstuhl 17, 67117 Heidelberg, Germany
         \and
         INAF–Istituto di Radioastronomia, Via Gobetti 101, 40129 Bologna, Italy
         \and
         European Southern Observatory (ESO), Karl-Schwarzschild-Straße 2, 85748 Garching bei München, Germany}
   \date{Received XXX; accepted XXX}

  \abstract{Water is one of the central molecules for the formation and habitability of planets. In particular, the region where water freezes-out, the water snowline, could be a favorable location to form planets in protoplanetary disks.}
  {We aim to spatially resolve the water emission in the HL~Tau disk using high resolution ALMA observations of the \ce{H2O} 183~GHz line ($E_u = 205$~K). We compare the spatially resolved \ce{H2O} emission with that of \ce{H^13CO+}, a chemical tracer of the water snowline to observationally test their anti-correlation. In addition, we aim to quantify the fraction of the water reservoir hidden by optically thick dust at ALMA wavelengths versus far- and mid-IR wavelengths.}
  {We use high resolution ALMA observations to spatially resolve the \ce{H2O} $3_{1,3}-2_{2,0}$ line at 183~GHz, \ce{H^13CO+} $J=2-1$, and the SO $4_4-3_3$ transition in the HL Tau disk. A rotational diagram analysis is used to characterize the water reservoir seen with ALMA and compare this to the reservoir visible at mid- and far-IR wavelengths. }
  {We find that the \ce{H2O} 183~GHz line has a compact central component and a diffuse component that is seen out to $\sim 75$~au. A radially resolved rotational diagram shows that the excitation temperature of the water is $\sim 350$~K independent of radius. The steep drop in the water brightness temperature outside the central beam of the observations where the emission is optically thick is consistent with the water snowline being located inside the central beam ($\lesssim 6$~au) at the height probed by the observations. Comparing the ALMA lines to those seen at shorter wavelengths shows that only $0.02\%-2\%$ of the water reservoir is visible at mid- and far-IR wavelengths, respectively, due to optically thick dust hiding the emission whereas 35-70\% is visible with ALMA. An anti-correlation between the \ce{H2O} and \ce{H^13CO+} emission is found but this is likely caused by optically thick dust hiding the \ce{H^13CO+} emission in the disk center. Finally, we see SO emission tracing the disk and for the first time in SO a molecular outflow and the infalling streamer out to $\sim 2"$. The velocity structure hints at a possible connection between the SO and the \ce{H2O} emission. }
  {Spatially resolved observations of \ce{H2O} lines at (sub-)mm wavelengths provide valuable constraints on the location of the water snowline, while probing the bulk of the gas-phase reservoirs.}


   \maketitle
%

\section{Introduction}

Water is a key molecule in shaping our Solar System. On a very local scale, water is essential for the life on Earth we know today. On the scale of (our) planetary system, the water snowline where water freezes out in the protoplanetary disk could have been a favorable location to start forming planets \citep{Drazkowska2017, Schoonenberg2017}. In addition, water is one of the main carriers of oxygen. Therefore, there is a major shift in the oxygen reservoir from gas to ice across the water snowline, changing the chemical composition of the planet-forming material \citep{Oberg2011, Oberg2021}. 

Despite the importance of the water snowline, locating it in a protoplanetary disk is difficult. In a typical disk around a T~Tauri or Herbig star, the snowline is located at a radius of a few up to $\sim 10$~au from the central star \citep{Harsono2015}. Thus, ground-based observatories, such as the Atacama Large Millimeter/submillimeter Array (ALMA) that can reach the required angular resolution, have to look through the Earth's atmosphere where gas-phase water can absorb the water emission from a protoplanetary disk or other astronomical objects.  

Space based-observatories such as \textit{Spitzer}, \textit{Herschel}, and JWST are not limited by the water in Earth atmosphere but lack the spatial resolution to resolve the water snowline. Still, some constraints on the water snowline location have been obtained by analyzing spectrally resolved line profiles \citep[e.g., ][]{Pontoppidan2010, vanDishoeck2021, Pirovano2022, Banzatti2025}. 
In addition, the water emission can be characterized through a rotational diagram analysis  or with thermochemical models \citep[e.g.,][]{Pirovano2022, Pontoppidan2024, Banzatti2023, Gasman2023, Temmink2024, Kaeufer2024, Vlasblom2025}. These methods can be used to distinguish the thermally desorbed water reservoir at a temperature of 150~K inside the water snowline from the cold photodesorbed reservoir in the outer disk below the sublimation temperature of water, and the reformed water in the disk surface layers where the gas temperature exceeds 300~K such that OH can react with \ce{H2} to form \ce{H2O}. 

To circumvent the Earth atmosphere and probe closer to the disk midplane, chemical tracers of the water snowline have been used with e.g., ALMA. From a chemical point of view, \ce{HCO+}, or the more optically thin \ce{H^13CO+} isotopologue, is one of the best tracers of the water snowline since it is very efficiently destroyed by gas-phase \ce{H2O} \citep{Phillips1992, Bergin1998}. Therefore, no \ce{HCO+} emission is expected in the disk center and a ring of emission is seen outside the water snowline where \ce{HCO+} can survive \citep{Leemker2021}. Still, optically thick dust complicates the analysis of \ce{HCO+} observations as this can also cause a central hole in the molecular line emission \citep{Isella2016, Weaver2018}. Therefore, \ce{HCO+} isotopologues are best used in disks with optically thin dust around the water snowline \citep{Leemker2021}. 

Recently, the water snowline has been located directly through emission of \ce{H2O} or its isotopologues in three bright disks. Spatially resolved emission of HDO and H$_2^{18}$O in the disk around the young, outbursting V883~Ori star drops off at 80~au, confirming the results from other chemical tracers and detailed modelling \citep{vantHoff2018, Leemker2021, Tobin2023, Wang2025}. The main water isotopologue has been used in the disk around the highly accreting AS~205 star where the water snowline was found to be vertical at a radial distance of 2 au by modelling the line profile of a high excitation line (1861.3~K upper energy level; \citealt{Carr2018, Bosman2021}).
In the HD~100546 disk, the snowline was found to be at the dust cavity wall \citep{Rampinelliinprep}. 

ALMA has the potential to resolve the water snowline in quiescent disks without a central cavity using the \ce{H2O} 183~GHz line. This line has an upper energy level comparable to the sublimation temperature of water ($E_u = 205$~K), thus probing the thermally desorbed water reservoir. \citet{Facchini2024} published the first, and to date only, detection of this line in a full and quiescent disk: HL~Tau. In this work, we combine these data with new high resolution observations of this line. 
HL~Tau is a massive Class~I/II protoplanetary disk of 0.2~$M_{\odot}$ surrounded by a protostellar envelope at a distance of 140~pc \citep{Rebull2004,  Robitaille2007, Furlan2008, Galli2018, Booth2020}. In addition, a streamer is seen in \ce{HCO+} that impacts the disk liberating SO and \ce{SO2} \citep{Yen2019, Garufi2022} although no \ce{CH3OH} is detected \citep{Soaveinprep}. In the mid- and far infrared (IR), hot gas-phase water is detected \citep{RiviereMarichalar2012, AlonsoMartinez2017, Salyk2019} that likely originates from the inner regions of the highly structured dust disk \citep{ALMA2015, CarrascoGonzalez2019, GuerraAlvarado2024}.

In Sect.~\ref{sec:obs}, the self-calibration of these data are described together with the \ce{H^13CO+} and SO line that are detected in the same dataset. Our results are presented in Sect.~\ref{sec:results}, where we observationally test the \ce{H2O}-\ce{HCO+} anti-correlation due to the water snowline. In addition, we compare the water reservoir visible with ALMA at (sub-)mm wavelengths to that in the mid- and far-IR. In Sect.~\ref{sec:disc} the implications for the water snowline location are discussed together with a comparison of the morphology of the \ce{H2O} and SO emission. Our conclusions are summarized in Sect.~\ref{sec:concl}.

\section{Observations} \label{sec:obs}

The main water isotopolgue has been targeted in the HL~Tau disk by two separate ALMA projects. In this work, we present new, high spatial resolution data of the \ce{H2O} line at 183~GHz in ALMA Band~5 that was observed as part of ALMA project 2022.1.00905.S (PI: S.~Facchini, \citealt{Facchini2024}). These data were combined with ALMA program 2017.1.01178.S (PI: E.~Humphreys, \citealt{Facchini2024}) that was taken in a more compact configuration to increase the maximum recoverable scale of the combined dataset. In addition, we reimaged the self-calibrated \ce{H2O} 321~GHz line in ALMA Band~7 observed by the latter program to allow for a uniform analysis with the 183~GHz line. The calibration of these data is presented in \citet{Facchini2024}.

\begin{table*}
\caption{ALMA observations covering the \ce{H2O} line at 183~GHz. \label{tab:obs_details}}
\centering
\begin{tabular}{lcccccccc}
\hline\hline
ALMA project & EB & baselines & integration  & pwv  & bandpass \& flux  & phase  & number of & First published \\
&  & (m) & time (h) & (mm) & calibrator & calibrator & antennas & in  \\
\hline
2017.1.01178.S & SB 0 & 15-1398  & 1.6  & 0.2       & J0423-0120 & J0510+1800 & 43 & \citealt{Facchini2024} \\
2022.1.00905.S & LB 0 & 85-8548 & 1.7 & 0.4 & J0522-3627 & J0431+1731 & 41 & this work \\
2022.1.00905.S & LB 1 & 113-9743 & 1.7 & 0.4 & J0423-0120 & J0431+1731 & 44 & this work \\ 
2022.1.00905.S & LB 2 & 83-11886 & 1.7 & 0.5 & J0522-3627 & J0431+1731 & 42 & this work \\ 
2022.1.00905.S & LB 3 & 92-8548  & 1.7 & 0.5 & J0423-0120 & J0431+1731 & 44 & this work \\ 
2022.1.00905.S & LB 4 & 92-8548  & 1.7 & 0.5 & J0423-0120 & J0431+1731 & 48 & this work \\ 
2022.1.00905.S & LB 5 & 92-8548  & 1.7 & 0.5 & J0423-0120 & J0431+1731 & 48 & this work \\ 
2022.1.00905.S & LB 6 & 92-8548  & 1.7 & 0.5 & J0423-0120 & J0431+1731 & 48 & this work \\ 
\hline
\end{tabular}
\end{table*}

HL~Tau was observed in the compact configuration for 1.6~hours, corresponding to a single execution block (EB). The spectral setup included a spectral window (spw) centered at the \ce{p-H2O} $3_{13}-2_{20}$ transition at 183.310~GHz with a spectral resolution of 122~kHz (0.2~km~s$^{-1}$). In addition, a continuum spw centered at 170.004~GHz with a spectral resolution of 977~kHz (1.7~km~s$^{-1}$) and 4 spws centered at 172.117~GHz, 172.709~GHz, 181.334~GHz, and 182.445~GHz with a spectral resolution of 122~kHz (0.2~km~s$^{-1}$) were included. 

The long baseline data (2022.1.00905.S) consist of 7~EBs of 1.7~hours each including calibration. These data cover the same \ce{H2O} 183~GHz line at the same spectral resolution but also included a spw with a resolution of 122~kHz (0.2~km~s$^{-1}$) centered at the \ce{H^13CO+} $J=2-1$ transition at 173.507~GHz ($E_u = 12.5$~K). In addition, the SO $4_4-3_3$ transition at 172.181~GHz ($E_u = 33.7$~K) is covered in the continuum spw centered at 171.525~GHz (977~kHz or 1.7~km~s$^{-1}$ spectral resolution). A second continuum spw was centered at 185.525~GHz. An overview of both programs is presented in Table~\ref{tab:obs_details}. 

We redid the self-calibration of the data covering the \ce{H2O} 183~GHz in the compact configuration (2017.1.01178.S) and we jointly self-calibrated these with the data taken in the extended configuration (2022.1.00905.S) following the methods of the exoALMA large program \citep{Loomis2025} using modules by \citet{Andrews2018, Czekala2021}. We used \texttt{CASA} version 6.5.4 \citep{McMullin2007, casa2022}. Before starting the self-calibration, the data within 15~km~s$^{-1}$ of the \ce{H2O} 183~GHz line and other bright lines were flagged after visual inspection of the data. This resulted in a total continuum bandwidth of 2.4~GHz for the short baseline and 4.2~GHz for the long baseline data. In addition, a phase-shift was applied to the high resolution data to align it with the low resolution data using the \texttt{phaseshift} task in \texttt{CASA}. The coordinate system was fixed to the new phase center using the task \texttt{fixplanets}. 

A single round of phase-only self-calibration was applied to each EB of the short and long baseline data separately. A model was created by cleaning the emission in each EB down to a 6$\sigma$ threshold using the task \texttt{tclean}. The phase-only gain solutions were found using the \texttt{gaincal} task where spws, scans, and polarizations were combined to a solution interval with the length of a single EB. In addition, only the antennas with at least 3 baselines and gain solutions with a signal-to-noise ratio of at least 4 were included. These solutions were applied to the data with \texttt{applycal} in the 'calonly' mode. All long-baseline EBs were then aligned with respect to the EB that was observed with the highest signal-to-noise ratio using the exoALMA alignment procedure \citep{Loomis2025}. The short-baseline data were aligned with the 7 concatenated long baseline EBs. 

Before combining the short and long baseline data, the phases of the short baseline data were self-calibrated through 5 subsequent rounds with decreasing solution intervals with the length of a single EB, 360, 120, 60, and 20 seconds. During the first round separate solutions were found for both polarizations whereas in the subsequent rounds the polarizations were combined to improve the signal-to-noise ratio on the solutions. This improved the peak signal-to-noise ratio on the continuum from 955 to 2682.

\begin{figure*}
\centering
\includegraphics[width=\hsize]{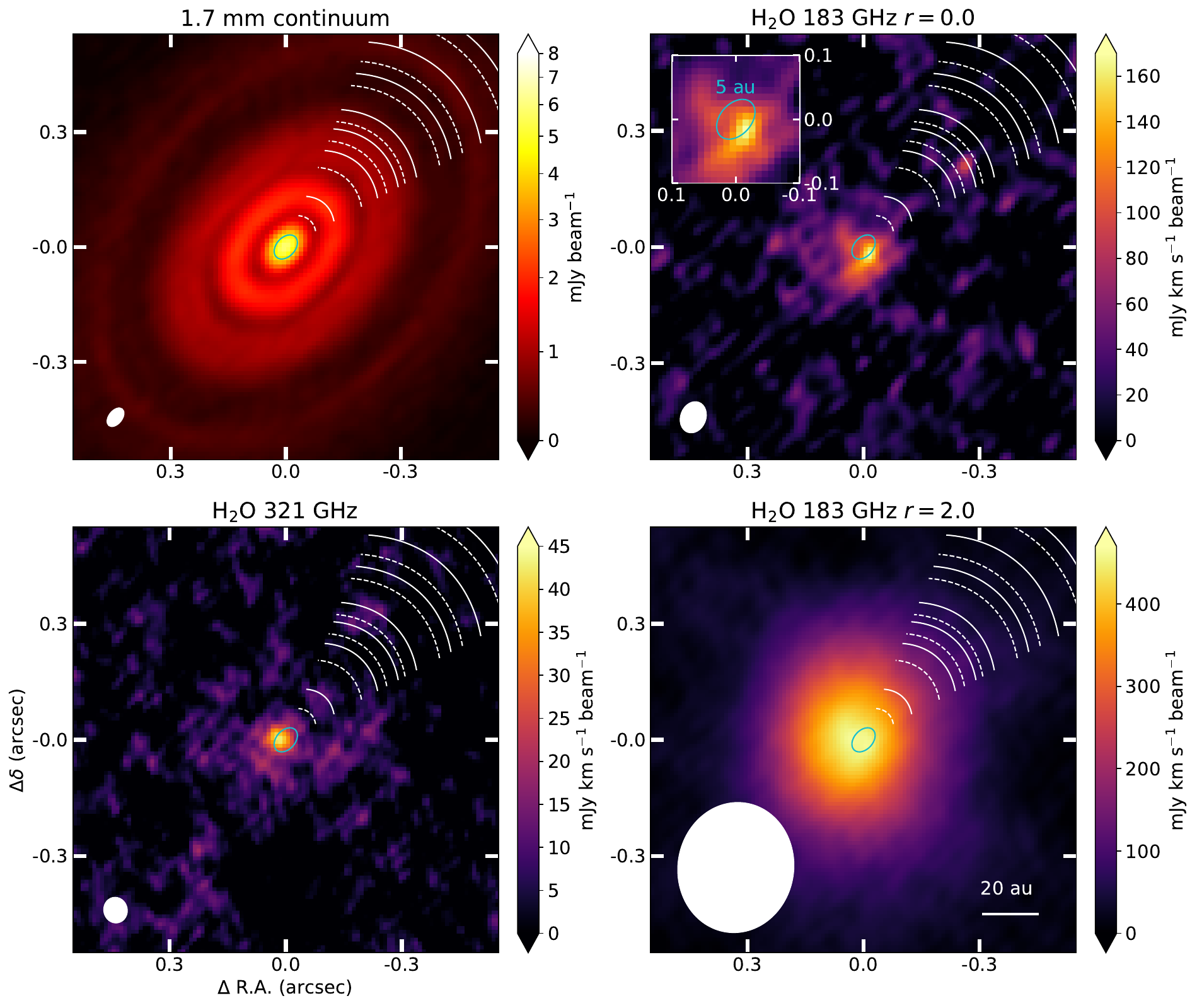}
  \caption{ALMA Band~5 images of the HL Tau disk and the 321~GHz \ce{H2O} line. The continuum (top left) and the JvM-corrected integrated intensity maps of the \ce{H2O} line at 183~GHz imaged with $r=0.0$ (top right) and $r=2.0$ (bottom right) providing high and moderate spatial resolution. The inset in the top right panel presents a zoom of the inner $0\farcs2$ and highlights the approximate water snowline location derived by \citet{GuerraAlvarado2024} The bottom left image is the reimaged \ce{H2O} 321~GHz line originally presented in \citet{Facchini2024}. The beams are indicated in the bottom left corners of the respective panels. The dust rings and gaps derived from high resolution ALMA observations are indicated with solid and dashed arcs in each panel \citep{ALMA2015, CarrascoGonzalez2019, GuerraAlvarado2024} and a 20~au scalebar is shown in the bottom right corner of the bottom right panel. }
     \label{fig:mom0_H2O}
\end{figure*}

The resulting self-calibrated short baseline data were combined with the long baseline data for a single round of phase-only joint self-calibration. The solution interval for this round was set to the length of a single EB while combining both spws and scans, and separate solutions were found for both polarizations. Additional rounds of self-calibration did not lead to an improvement in the signal-to-noise ratio of the resulting continuum image. Comparing the total flux revealed that the flux measured in all EBs are within 4\% of that in LB EB3 except for LB EB 2. Following \citet{Loomis2025}, we rescale the flux in this EB and repeat the joint short and long baseline self-calibration. 

Finally, we performed two rounds of amplitude and phase-self-calibration on the data. The models for these rounds of self-calibration were obtained by cleaning the image down to a 1$\sigma$ threshold to include as much flux as possible in the model. Then solutions were found for EB- and scan-long intervals by combining spws, scans, and polarizations and by combining solely spws and polarizations, respectively. This increased the peak signal to noise ratio on the continuum from 1102 after the SB self-calibration but before the joint self-calibration to 1365 after the joint self-calibration.

The gain solutions of the self-calibration were applied to the data containing the lines following the same order of the procedures as during the self-calibration. The data were continuum subtracted using the \texttt{CASA} task \texttt{uvcontsub\_old} by fitting a first order polynomial to the line free channels.

\begin{table*}
\caption{Observed molecular lines with ALMA in the HL~Tau disk.\label{tab:det_ALMA}}
\centering
\begin{tabular}{p{0.24\columnwidth}p{0.12\columnwidth}p{0.12\columnwidth}p{0.18\columnwidth}p{0.1\columnwidth}p{0.1\columnwidth}p{0.1\columnwidth}p{0.37\columnwidth}p{0.05\columnwidth}p{0.12\columnwidth}} 
\hline\hline
Transition & Freq. & $r_{\rm flux}$ & Int. flux & $\Delta V_{\rm chan}$ &$\sigma_{\rm chan}$ & \texttt{robust} & Beam & $\epsilon $& Baselines  \\
  & (GHz)& & (mJy km~s$^{-1}$) & (km~s$^{-1}$) & (mJy beam$^{-1}$) & &  &  &   \\
\hline
\ce{H2O} $3_{1,3}-2_{2,0}$  & 183.310 & $0\farcs28$ & $882 \pm 318$ & 1.0 &9-11& 0.0 & $0\farcs083 \times 0\farcs066\ (-21.0\degree)$ & 0.35 & SB, LB \\
\ce{H2O} $3_{1,3}-2_{2,0}$  & 183.310 & $0\farcs7$  & $906 \pm 140$ & 1.0 & 5-7 & 2.0 & $0\farcs35 \times 0\farcs30\ (-8.5\degree)   $ & 0.34 & SB, LB \\
\ce{H2O} $10_{2,9}-9_{3,6}$ & 321.226 & $0\farcs28$ & $269 \pm 91$ & 1.0 & 2 & 2.0 & $0\farcs067 \times 0\farcs061\ (12.3\degree)  $ & 0.39 & LB \\ 
\ce{H^13CO+} $2-1$ & 173.507 & $1\farcs2$ & $113 \pm 16^{(a)}$ & 1.0 & 0.7 & 2.0$^{(b)}$ & $0\farcs29\times 0\farcs26\ (-50.5\degree)$ & 0.86 & LB \\ 
\ce{SO}$^{(c)}$  $4_4-3_3$ & 172.181 & $0\farcs17$ & $106 \pm 13$ & 2.0 & 0.3 & 0.5 & $0\farcs11\times 0\farcs086\ (-33.1\degree)$ & 0.68 & LB \\
\ce{SO}$^{(d)}$  $4_4-3_3$ & 172.181 & $2\farcs5$ & $1019 \pm 155$ & 2.0 & 0.3 & 0.5 & $0\farcs11\times 0\farcs086\ (-33.1\degree)$ & 0.68 & LB \\
\hline
\end{tabular}
\tablefoot{The channel rms for the H2O 183~GHz transition is given as a range because the rms is frequency dependent due to the telluric as detailed in Appendix~\ref{app:telluric}. The uncertainty on the disk integrated flux includes the 10\% absolute flux cal error of ALMA. $^{(a)}$~Measured in the Keplerian mask; $^{(b)}$~a $0\farcs2$-$uv$ taper was applied; $^{(c)}$ tracing primarily the SO emission in the disk; $^{(d)}$ tracing the SO emission in the disk, outflow, and streamer.}
\end{table*}

\subsection{Imaging}

The data were imaged with the \texttt{CASA} task \texttt{tclean} using Briggs weighting. As the continuum is seen at a very high peak signal-to-noise ratio of 1365 when imaged with a \texttt{robust} parameter of 0.5 used for the self-calibration, the final continuum image is made with a \texttt{robust} parameter of $-0.5$, providing a high spatial resolution of $0\farcs056\times 0\farcs035\ (-38.7\degree)$ without losing too much in the signal-to-noise ratio. The continuum flux density within a $1\farcs0$ radius circular region centered on the continuum peak is $335 \pm 34$~mJy, where the uncertainty is dominated by the absolute flux calibration uncertainty of ALMA.

The water line is weaker and therefore imaged with a \texttt{robust} parameter of 0.0 and 2.0 and a channel width of 1~km~s$^{-1}$. The former provides a very good spatial resolution of $0\farcs083 \times 0\farcs066\ (-21.0\degree)$ at the cost of a larger channel rms noise. The latter weighting scheme provides a much larger beam of $0\farcs35 \times 0\farcs30\ (-8.5\degree)$, increasing the sensitivity to the diffuse \ce{H2O} emission at larger spatial scales (see the top right and bottom right panel in Fig.~\ref{fig:mom0_H2O}). The channel rms of these cubes is frequency dependent due to telluric around the targeted \ce{H2O} line. Therefore, the noise in each channel depends on the position of that channel with respect to the telluric at the time of the observations. We refer the reader to Appendix~\ref{app:telluric} and \citet{Rampinelliinprep} for a detailed discussion. The \ce{H2O} channel maps are presented in Fig.~\ref{fig:chans_H2O_r0.0} and \ref{fig:chans_H2O_r2.0} where the those in the former figure are imaged at a spectral resolution of 5 km~s$^{-1}$ to increase the signal-to-noise ratio in the channel maps.

The \ce{H^13CO+} was only covered by the long baseline data. The \ce{H^13CO+} $J=2-1$ line ($E_u$ = 12.5~K) was imaged in 1~km~s$^{-1}$ wide channels with a \texttt{robust} parameter of 2.0 and a $0\farcs2\ uv$-taper to increase the sensitivity to extended emission. This resulted in a cube with a channel rms of 0.7~mJy~beam$^{-1}$. 

The SO $4_4-3_3$ line ($E_u = 33.7$~K) was covered by the continuum spw in the long baseline data only, and was imaged with a \texttt{robust} parameter of 0.5 and a channel width of 2~km~s$^{-1}$. As this line is far from the center of the telluric, the resulting channel rms of the SO cube is only 0.3~mJy~beam$^{-1}$. 

Finally, we reimaged the \ce{H2O} line at 321~GHz using the self-calibrated ms tables of \citet{Facchini2024} to allow for a uniform analysis of the data. The continuum subtracted image was recreated following \citet{Facchini2024} but with a 1~km~s$^{-1}$ channel width, resulting in a channel rms of 2~mJy~beam$^{-1}$. Non-continuum subtracted images to measure the brightness temperature of the \ce{H2O} 183~GHz and 321~GHz lines were made following the same procedure that was used for the continuum subtracted images.

\subsection{JvM-correction} \label{sec:JvM}
In order to accurately retrieve line fluxes and channel map intensities, we apply the so-called JvM correction to the line cubes \citep{JvM1995}. As detailed in \citet{Czekala2021}, this is needed in order not to overestimate line fluxes for extended and low signal-to-noise ratio emission morphologies, where a large fraction of the emission is not included in the \texttt{clean} model. In our paper, this is particularly important for the water line flux measurements. We verified that this correction is needed to maintain the uniform line fluxes across a wide range of \texttt{robust} parameters used in \texttt{tclean}. We note that \citet{Facchini2024} did not apply the JvM correction, thus overestimating the 321~GHz water line flux. The 183~GHz water line in \citet{Facchini2024} is significantly less affected due to the higher signal-to-noise ratio in the SB data. Since the JvM correction is known to underestimate the point source rms \citep{casassus2022}, uncertainties on flux measurements are estimated on the non-JvM-corrected images (Section~\ref{sec:uncertainties}). The so-called $\epsilon$ factors (used to rescale the residuals in the \texttt{clean} image restoration) are reported in Table~\ref{tab:det_ALMA}.

\section{Methods}

To further analyze the data, the imagecubes are collapsed to moment maps and spectra as described below. In addition, the rotational diagram analysis to derive column densities is briefly reiterated. 

\subsection{Moment maps and integrated fluxes}
The moment 0, 1, and 8 maps are computed using the \texttt{Better Moments} package \citep{Teague2019}. No clip was applied to the integrated and peak intensity maps (moment 0 and 8) and a 4$\sigma$ clip was applied to the cubes before obtaining the moment 1 maps. In addition, a Keplerian mask was applied to the \ce{H^13CO+} cube before making the integrated intensity map as this line was found to generally follow a Keplerian pattern (see Fig.~\ref{fig:mom01_H13CO+}). This mask predicts where the \ce{H^13CO+} emission emits in the channel maps based on the disk inclination of 46.7$\degree$, the position angle of 138$\degree$, a distance of 140~pc, a stellar mass of 2.1~M$_{\odot}$, and the source velocity of 7.1~km~s$^{-1}$ \citep{Rebull2004, ALMA2015, Galli2018, Yen2019, Garufi2021, Garufi2022}. The outer radius of the mask is set to a large radius of 20" to include all flux. Finally, the mask is convolved with a Gaussian 1.5 times larger than the beam. 

The maps of the \ce{H2O} lines are created by integrating the cubes from -8.5 to 22.5~km~s$^{-1}$ for the 183~GHz line and from -10.4 to 24.6~km~s$^{-1}$ for the 321~GHz line. The latter velocity range is identical to that in \citet{Facchini2024} but the former is larger as the 183~GHz line is detected out to higher velocity offsets from the systemic velocity of 7.1~km~s$^{-1}$ with the addition of the long baseline data. The \ce{H^13CO+} and SO emission are integrated over a range from -8.5 to 22.5~km~s$^{-1}$ and from -5 to 19~km~s$^{-1}$, respectively.

The disk integrated fluxes were computed using the same velocity limits as were used for the moment maps. We note that the \ce{H^13CO+} and the SO emission in the channel maps extend to scales up to or beyond the maximum recoverable scale of $1\farcs7$ of the long baseline data (see Fig.~\ref{fig:chans_H13CO+} and \ref{fig:chans_SO}). Therefore, some flux might be resolved out. An overview of the detected lines is presented in Table~\ref{tab:det_ALMA}.

\subsection{Uncertainties on the disk integrated flux and azimuthally averaged radial profiles}
\label{sec:uncertainties}

The uncertainty on the disk integrated flux is estimated by propagating the channel rms from the non-JvM-corrected cube:
\begin{align}
    \sigma_{F_{\nu}\Delta V} = \sigma_{\rm chan} \times \Delta V_{\rm chan} \times \sqrt{N_{\rm pix\ mask}/N_{\rm pix\ beam}},
\end{align}
with $\sigma_{\rm chan}$ the rms in the channels, $\Delta V_{\rm chan}$ the channel width, $N_{\rm pix\ mask}$ the total number of pixels in space and velocity used to calculate the integrated flux $F_{\nu}\Delta V$, and $N_{\rm pix\ beam}$ the number of pixels per beam. The 10\% absolute flux calibration uncertainty of ALMA is added in quadrature. 

Similarly, the uncertainty on the azimuthally averaged radial profiles obtained using the gofish package \citep{Teague2019} are estimated from the channel rms. First, the uncertainty map of the integrated intensity is estimated following e.g., \citet{Leemker2023}:
\begin{align}
    \sigma_{\rm mom0} = \sigma_{\rm chan} \times \Delta V_{\rm chan} \times \sqrt{N_{\rm chan}},
\end{align}
with $N_{\rm chan}$ the number of channels the integrated intensity map is computed over. In case a Keplerian mask is used, $N_{\rm chan}$ depends on the location in the map with positions close to the star generally having more channels than those farther away. The uncertainty in each bin of the azimuthally averaged radial profile then follows:
\begin{align}
    \sigma_{\rm radial\ profile} = \sqrt{ \frac{1}{N_{\rm beams\ per\ bin}N_{\rm pix\ per\ bin}} \sum_{\rm pix} \sigma^2_{\rm mom0} },
\end{align}
with $N_{\rm beams\ per\ bin}$ the number of beams per bin and $N_{\rm pix\ per\ bin}$ the number of pixels in each radial bin. As the bins have a width of a quarter of the beam major axis, the number of beams per bin is calculated as the ratio of the circumference of the ellipse that defines the radial bin and the average of the beam major and minor axes.

\subsection{Rotational diagram}

The resulting \ce{H2O} 183~GHz line is imaged at a similar spatial scale to the 321~GHz line originally presented in \citet{Facchini2024}. As these lines have different upper energy levels, the ratio of their observed flux is a probe of their excitation temperature when convolved to the same beam size. As the 321~GHz line is observed at a slightly higher spatial resolution, this cube is convolved to match the beam size of the 183~GHz line. 

The formalism by \citet{Goldsmith1999} and \citet{Loomis2018} is briefly summarized below. If the emission is optically thin, the integrated flux is directly related to the column density of the upper level $u$ through:
\begin{align}
    N_u^{\rm thin} = \frac{4\pi F_{\nu}\Delta V}{A_{ul}\Omega hc},    
\end{align}
with $A_{ul}$ the Einstein-$A$ coefficient for spontaneous emission, $\Omega$ the assumed emitting region, $h$ the Planck constant, and $c$ the speed of light. 
For the radially resolved rotational diagram, the emitting area is assumed to be the beam size.

\begin{figure*}
\centering
    \begin{subfigure}[b]{0.49\textwidth}
    \centering
    \includegraphics[width=\textwidth]{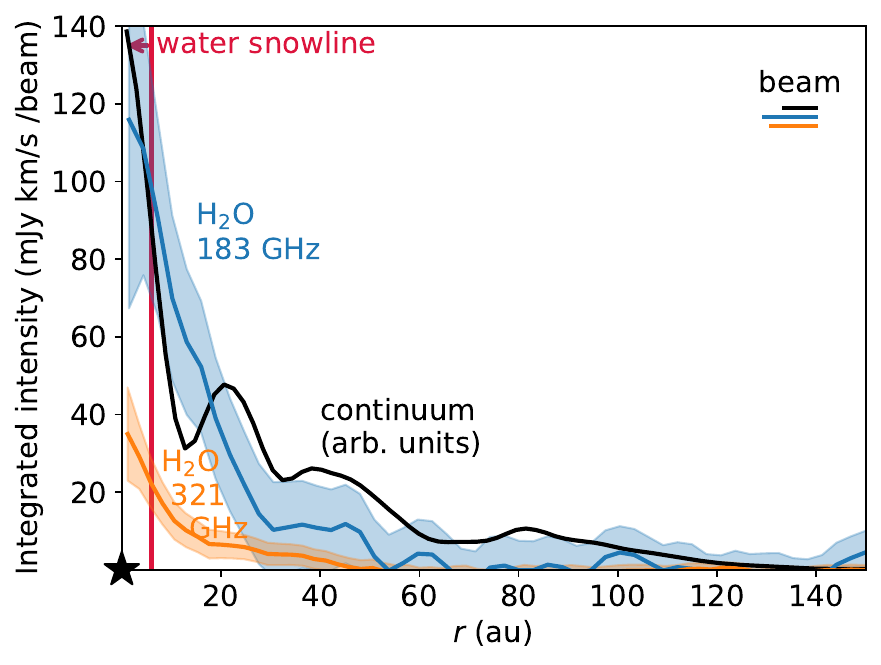}
    \end{subfigure}
    \begin{subfigure}[b]{0.49\textwidth}
    \centering
    \includegraphics[width=\textwidth]{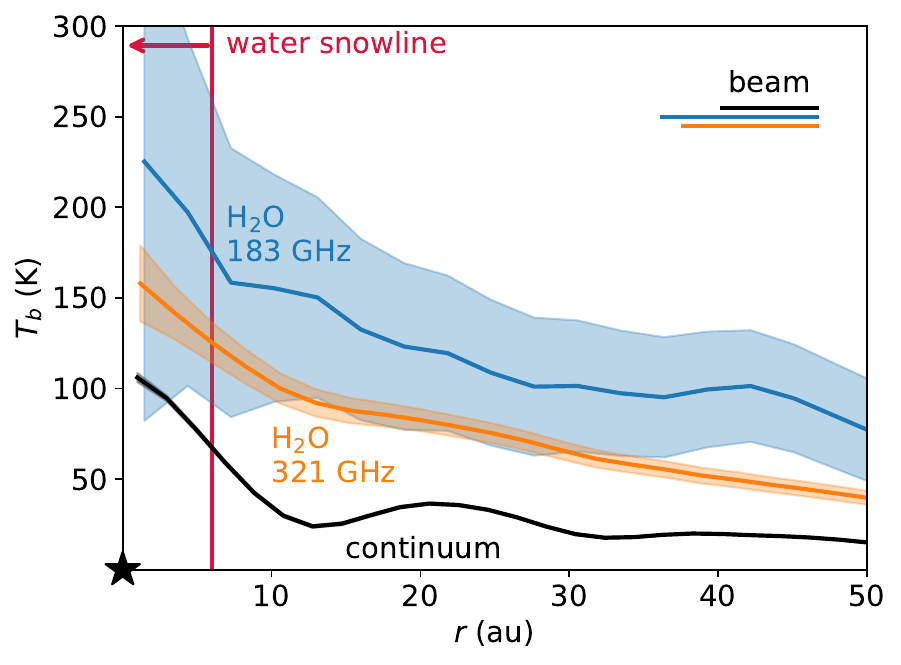}
    \end{subfigure}
    \caption{Azimuthally averaged radial profiles of the integrated intensity (left) and brightness temperature (right). The \ce{H2O} line at 183~GHz imaged with a \texttt{robust} parameter of 0.0 is indicated in blue, the \ce{H2O} line at 321~GHz in orange, and the continuum emission in black. In the left panel the continuum emission is presented in arbitrary units and the upper limit on the water snowline location is indicated with the vertical red line. Note the difference in the radial axes. The beams are indicated with the horizontal bars in the top right corner.} 
    \label{fig:azi_avg_H2O}
\end{figure*}

For (marginally) optically thick lines, the assumption of optically thin emission leads to an underestimation of the column density of level $u$ by a factor of $C(\tau_{\nu})$:
\begin{align}
    C(\tau_{\nu}) = \frac{\tau_{\nu}}{1-e^{-\tau_{\nu}}} \\
    N_u = N_u^{\rm thin} C(\tau_{\nu}),    
\end{align}    
with $\tau_{\nu}$ the optical depth of the line. This can be approximated as:
\begin{align}
    \tau_{\nu} = \frac{A_{ul}c^3N_u}{8\pi \nu^3 \Delta V} (e^{h\nu/k_BT}-1 ), \label{eq:tau}
\end{align}
with $\nu$ the frequency of the line, $k_{\rm B}$ the Boltzmann constant, $T$ the excitation temperature, and $\Delta V = \sqrt{8k_BT\ln(2)/m}$ the thermal line width for a molecule of mass $m$. 

Under the assumption of local thermodynamical equilibrium (LTE), the level populations follow a Boltzmann distribution:
\begin{align}
    \frac{N_u}{g_u} = \frac{N_{\rm tot}}{Q(T)} e^{-E_u/k_{\rm B}T}, \label{eq:rot_dia}
\end{align}
with $g_u$ and $E_u$ the degeneracy and energy of level $u$, respectively, $N_{\rm tot}$ the total water column density, and $Q(T)$ the partition function. An overview of the disk integrated line fluxes is presented in Table~\ref{tab:det_ALMA} for the lines observed with ALMA and in Table~\ref{tab:all_H2O} for all \ce{H2O} lines seen in the HL~Tau disk with ALMA, \textit{Herschel}, and Gemini North. 
In addition, the latter table includes the line constants for the \ce{H2O} transitions analyzed in this work. Those for the lines observed with ALMA and \textit{Herschel} are taken from the JPL database \citep{Pickett1998, Yu2012} whereas those for the high excitation lines seen with Gemini North are not included in the JPL database. Therefore, the latter are taken from the LAMDA database \citep{Tennyson2001, Schoier2005, Barber2006, Faure2008} where $g_u$ is corrected for the assumed ortho-to-para ratio of 3 used by the partition function in the JPL database and appropriate for \ce{H2O} that formed in the gas-phase or was thermally desorbed \citep{Cheng2022}.

The uncertainty on the temperature, column density, and optical depth is estimated from the uncertainty on the integrated flux including the 10\% absolute flux calibration uncertainty of ALMA as the \ce{H2O} lines are part of different ALMA projects and compared to lines seen with \textit{Herschel} and Gemini North. First, the uncertainty on the temperature is estimated. As the temperature cannot be isolated in Eq.~\ref{eq:rot_dia} due to the dependence in $N_u$, $Q(T)$, and $e^{-E_u/k_{\rm B}T}$, the equation is linearized through a first order Taylor polynomial around $\bar{T}$:
\begin{align}
    T \approx - f \frac{1}{\left. \diff{f}{T}\right |_{\bar{T}}} + \bar{T}, \label{eq:T_TE}
\end{align}
where $f$ is defined as
\begin{align}
    f \equiv -\frac{E_1-E_2}{T} - \ln(C_1) + ln(C_2) - \ln \left (\frac{N_1^{\rm thin} g_2}{N_2^{\rm thin} g_1}\right ),
\end{align}
with subscripts 1 and 2 referring to the two lines used for the rotational diagram analysis. The uncertainty on $T$ is then estimated using the standard propagation of uncertainty on Eq.~\ref{eq:T_TE} where the derivatives are calculated numerically over a 5\% step size in each parameter. The uncertainty on $N_{\rm tot}$ and $\tau_{\nu}$ are estimated using the standard propagation of uncertainty on Eq.~\ref{eq:tau} and \ref{eq:rot_dia} taking the error on the temperature and integrated flux into account.

\section{Results} \label{sec:results}

\subsection{\ce{H2O} morphology}

The self-calibrated continuum image and the integrated intensity maps of the \ce{H2O} 183~GHz and 321~GHz lines are presented in Fig.~\ref{fig:mom0_H2O}. The \ce{H2O} emission imaged with a small \texttt{robust} parameter of 0.0 providing high angular resolution and good sensitivity to point sources is presented in the middle panel. The emission is compact and peaks at a projected distance of 5~au in the south west, consistent with the center of the disk within the size of the beam of $0\farcs083\times 0\farcs066$ ($12\times9$~au). Following \citet{Fasano2025}, the significance of this offset can be estimated using the uncertainty on the astrometric accuracy $\theta_{\rm FWHM}/(0.9\times S/N)$ with $\theta_{\rm FWHM}$ the size of the beam and $S/N$ the signal-to-noise ratio of the peak that we conservatively measure using the peak integrated intensity in the JvM-corrected moment 0 map and the noise in the non-JvM corrected map. As the astrometric uncertainty of $0\farcs03$ is comparable to 5~au projected distance, the offset is not significant at the signal-to-noise ratio of the data. 

\begin{figure*}
\centering
\includegraphics[width=\hsize]{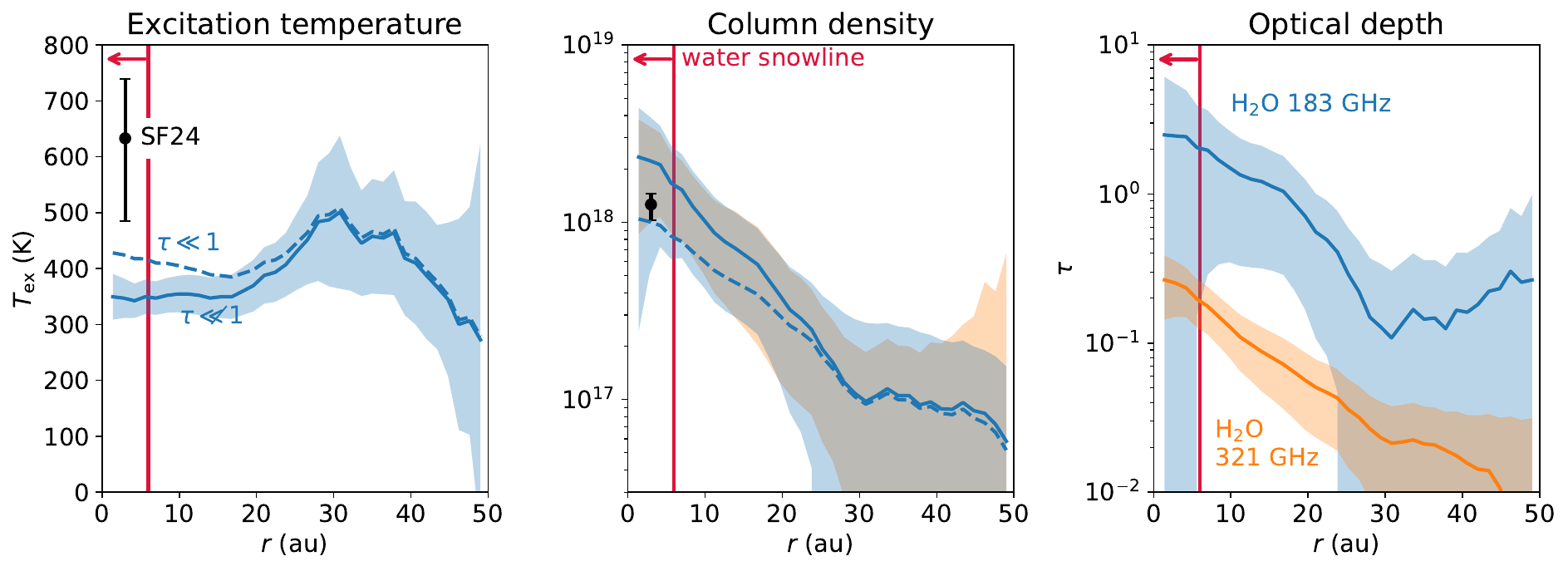}
  \caption{Radially resolved rotational diagram analysis of the \ce{H2O} emission in the HL~Tau disk. The excitation temperature (left), column density (middle), and optical depth (right) of the \ce{H2O} 183~GHz (blue) and 321~GHz (orange) lines are obtained after convolving both lines to the same beam size. The dashed lines are derived under the assumption of fully optically thin emission whereas the solid lines include the correction for the optical depth of the lines. The black scatter points indicate the disk averaged results from \citet{Facchini2024} assuming an emitting region of 17~au and the upper limit on the water snowline location is indicated with the vertical red line.}
     \label{fig:rot_dia_res}
\end{figure*}

The \ce{H2O} 183~GHz integrated intensity map imaged at a higher \texttt{robust} parameter of 2.0 is presented in the right hand panel. The water emission in this panel is seen out to much larger radii due to the better sensitivity to diffuse and extended emission. Still, the \ce{H2O} emission is more compact than the continuum disk. A slight asymmetry is seen in the \ce{H2O} emission when imaged at moderate angular resolution where the diffuse water emission extends somewhat towards the north or north east side of the disk.

The azimuthally averaged radial profiles of the high resolution \ce{H2O} 183~GHz and 321~GHz lines together with that of the continuum emission are presented in the left panel of Fig.~\ref{fig:azi_avg_H2O}. The \ce{H2O} 183~GHz line is centrally peaked with a steep drop off outwards. Imaging the \ce{H2O} 183~GHz line with higher \texttt{robust} parameters and thus larger beam sizes that are more sensitive to diffuse emission, reveals that the water emission extends out to $\sim 75$~au (see Fig.~\ref{fig:azi_avg_H2O_low_res} in Appendix~\ref{app:maps_and_spectra}). Similarly, the hot 321~GHz line shows a bright central component and an extended shoulder of emission between $\sim 10$ and 40~au. This suggests that the central component of the \ce{H2O} 183 and 321~GHz lines probes a large column of gas-phase \ce{H2O} inside the water snowline.

This is further supported by the brightness temperature profiles of the lines that are presented in the right panel of Fig.~\ref{fig:azi_avg_H2O} derived using the Rayleigh-Jeans approximation 
and the peak intensity map extracted from the image cube between $\mathrm{-8.5~km~s}^{-1}$ to 22.5~km~s$^{-1}$ without continuum subtraction and without shifting each pixel by the project Keplerian velocity at that location. Similar to the integrated intensity profiles, the shifting and stacking was not applied as this can underestimate the intensity in the inner disk regions where the water snowline is expected to be located. The brightness temperature profiles are only shown in the inner 50~au because the peak intensity in the outer disk regions is dominated by the noise in the data cubes as the peak flux is measured over a large velocity range. The \ce{H2O} 183~GHz and 321~GHz lines reach a high brightness temperature of 220~K and 160~K in the inner beam, respectively. These temperatures are a lower limit to the kinetic temperature of the gas due to the finite line optical depth, spatial, and spectral resolution of the data \citep{Leemker2022} and assume LTE emission. Therefore, the gas probed by the central beam of the observations is sufficiently warm for thermal sublimation of water inside the water snowline at $T \sim 150$~K. Outside the central beam, the brightness temperature steeply drops, suggesting that the water becomes less optically thick.

\subsection{Excitation of \ce{H2O} at sub-mm wavelengths}

To analyze the origin of the water in the HL Tau disk, we perform a rotational diagram analysis using the warm 183~GHz and the hot 321~GHz line with upper energy levels of 205~K and 1861~K, respectively. The integrated intensity map of the 321~GHz line is convolved to the beam size of the 183~GHz line to compare the lines on the same spatial scale. The results assuming LTE are presented in Fig.~\ref{fig:rot_dia_res} where the dashed lines assume optically thin emission and the solid lines present the results taking the optical depth into account. The column density derived from the 183 and 321~GHz lines are identical by construction though their uncertainties are not due to the dependence on the uncertainty on the excitation temperature and the line flux itself. The assumption of LTE is further discussed in Sect.~\ref{sec:IR}.

The excitation temperature of the \ce{H2O} line is $\sim350$~K and independent of radius within the uncertainty of the data. Following the results from the radial profiles, the rotational diagram can be divided into two regions: the inner beam where the 183~GHz line is optically thick and remaining disk regions. 
In the inner regions, the optical depth of the 183~GHz line is an order of magnitude higher than that of the 321~GHz line though the precise value is uncertain, see right panel Fig.~\ref{fig:rot_dia_res}. As the warm 183~GHz line is more optically thick than the hot 321~GHz line, the rotational temperature in the inner beam overestimates the kinetic temperature of the \ce{H2O} gas. Therefore, the inner beam likely probes a large column of gas-phase \ce{H2O} at a temperature of at most $350$~K, consistent with the thermally desorbed water reservoir. As the lines are (marginally) optically thick in this region, the observations likely do not probe the disk midplane. Thus, the water snowline is likely located inside the inner beam ($\lesssim 6$~au) at the height in the disk probed by the observations.

\begin{figure*}
\centering
\includegraphics[width=\hsize]{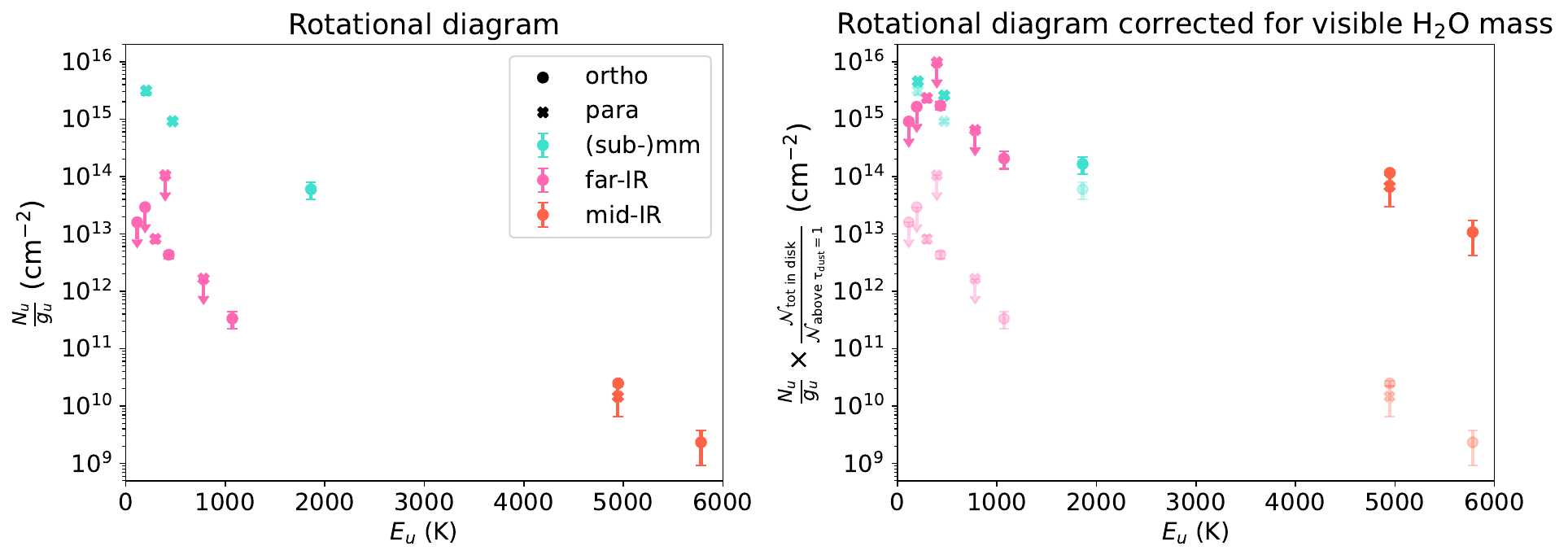}
  \caption{Rotational diagram summarizing the disk integrated \ce{H2O} line detections in the HL~Tau disk with ALMA in the (sub-)mm (cyan), \textit{Herschel} PACS in the far-IR (pink), and Gemini North in the mid-IR (orange). The left panel shows the rotational diagram without a correction for the difference in the continuum optical depth for the various lines whereas in the right panel this correction factor derived from a thermochemical model is applied. The shaded points in the right panel are identical to those in the left panel.}
     \label{fig:rot_dia_unres}
\end{figure*}

Outside the central beam, the emission is less optically thick as the brightness temperature  drops well below the excitation temperature. The excitation temperature of 350~K is consistent with the expected temperature of $300$~K above which water can reform from OH. Therefore, the emission is likely tracing the radially extended warm water reservoir in the disk surface. 

The column density and excitation temperature derived from the high resolution 183 and 321~GHz data presented in this work are roughly consistent with those derived from these lines and the 325~GHz line at lower spatial resolution in \citet{Facchini2024}. Their \ce{H2O} column density assuming that all emission originates from within 17~au is consistent with that in the inner beam within a factor of 2. The excitation temperatures are somewhat different due to the overestimation of the flux of the hot 321~GHz line (see Sect.~\ref{sec:JvM}) and the addition of the 325~GHz line ($E_u$ = 470~K) in \citet{Facchini2024} that is only available at low spatial resolution and thus not included in the analysis presented in this work.

\begin{figure*}
\centering
\includegraphics[width=\hsize]{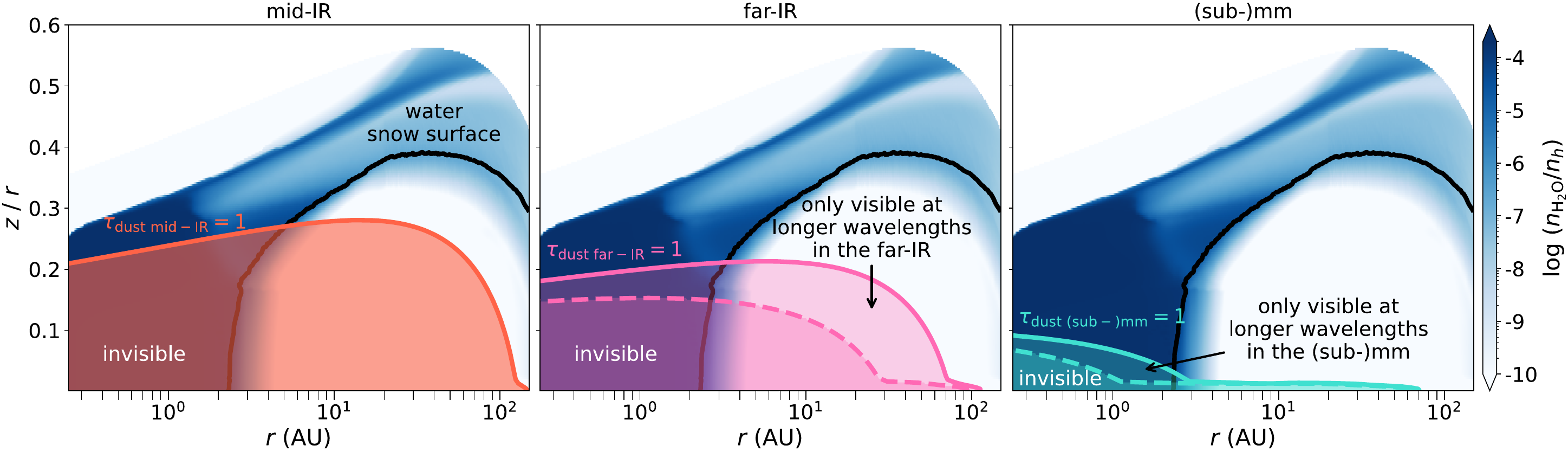}
  \caption{Abundance of gas-phase \ce{H2O} in a representative model for the HL~Tau disk (colored background). The orange, pink, and cyan contours in the left, middle, and right panels indicate the $\tau_{\rm dust}=1$ surface for the lines in Table~\ref{tab:all_H2O} in the mid-IR, far-IR, and (sub-)mm, respectively. In the middle and right panel the solid and dashed lines indicate the long and short wavelength ends of the range of lines in Table~\ref{tab:all_H2O} in the far-IR and (sub-)mm, respectively. The water snow surface in the model is indicated with the black contour.}
     \label{fig:multiplot_H2O}
\end{figure*}

\begin{table*}
\caption{Disk integrated \ce{H2O} line fluxes in the HL~Tau disk. \label{tab:all_H2O}}
\centering
\begin{tabular}{lcccccccccc}
\hline\hline
Molecule & Transition & Frequency  & Wavelength & Telescope & $\log(A_{ul})$ & $E_u$ & $g_u$ & Integrated flux & $\frac{\mathcal{N}_{\rm above\ \tau_{\rm dust} = 1}}{\mathcal{N}_{\rm tot\ in\ disk}}$ & Refs.   \\
 & & (GHz) & $(\mu m)$ &  & $\log(\rm{s^{-1}})$ & (K) &  & (Jy km~s$^{-1}$) & &    \\
\hline
\ce{p-H2O} & $3_{1,3}-2_{2,0}$ & 183.310 & 1635.44 & ALMA & -5.44 & 205 & 7 & 0.91 $\pm$ 0.14 & 68.90\% & 1,2\\
\ce{o-H2O} & $10_{2,9}-9_{3,6}$ & 321.226 & 933.28 & ALMA & -5.21 & 1861 & 63 & 0.27 $\pm$ 0.09 & 36.18\% & 1,2\\
\ce{p-H2O} & $5_{1,5}-4_{2,2}$ & 325.150 & 922.01 & ALMA & -4.94 & 470 & 11 & 1.33 $\pm$ 0.09 & 35.42\% & 1\\
\ce{o-H2O} & $2_{2,1}-2_{1,2}$ & 1661.01 & 180.49 & \textit{Herschel} & -1.52 & 194 & 15 & < 153 & 1.78\% & 3\\
\ce{o-H2O} & $2_{1,2}-1_{0,1}$ & 1669.90 & 179.53 & \textit{Herschel} & -1.26 & 114 & 15 & < 153 & 1.76\% & 3\\
\ce{p-H2O} & $4_{1,3}-3_{2,2}$ & 2074.43 & 144.52 & \textit{Herschel} & -1.48 & 396 & 9 & < 361 & 1.08\% & 3\\
\ce{p-H2O} & $3_{2,2}-2_{1,1}$ & 3331.46 & 89.99 & \textit{Herschel} & -0.46 & 297 & 7 & 229 $\pm$ 38 & 0.35\% & 3\\
\ce{p-H2O} & $6_{1,5}-5_{2,4}$ & 3798.28 & 78.93 & \textit{Herschel} & -0.35 & 781 & 13 & < 110 & 0.26\% & 3\\
\ce{o-H2O} & $4_{2,3}-3_{1,2}$ & 3807.26 & 78.74 & \textit{Herschel} & -0.32 & 432 & 27 & 646 $\pm$ 100 & 0.25\% & 3\\
\ce{o-H2O} & $8_{1,8}-7_{0,7}$ & 4734.30 & 63.32 & \textit{Herschel} & 0.24 & 1071 & 51 & 339 $\pm$ 115 & 0.16\% & 3,4\\
\ce{p-H2O} & $16_{3,13}-15_{2,14}$ & 24163.2 & 12.41 & GN & 0.63 & 4945 & 33 & 24 $\pm$ 13 & 0.02\% & 5\\
\ce{o-H2O} & $17_{4,13}-16_{3,14}$ & 24184.2 & 12.40 & GN & 0.89 & 5781 & 105 & 22 $\pm$ 13 & 0.02\% & 5\\
\ce{o-H2O} & $16_{4,13}-15_{1,14}$ & 24224.3 & 12.38 & GN & 0.63 & 4948 & 99 & 121 $\pm$ 13 & 0.02\% & 5\\
\hline
\end{tabular}
\tablefoot{The \ce{H2O} line constants of the lines observed with ALMA or \textit{Herschel} are taken from the JPL database \citep{Pickett1998, Yu2012}, and those observed with Gemini North are taken from the LAMDA database \citep{Tennyson2001, Schoier2005, Barber2006, Faure2008}. References: 1.~\citealt{Facchini2024}, 2.~this work, 3.~\citealt{AlonsoMartinez2017}, 4.~\citealt{RiviereMarichalar2012}, and 5.~\citealt{Salyk2019}}
\end{table*}

\subsection{\ce{H2O} hidden by optically thick dust} \label{sec:IR}
The \ce{H2O} 183 and 321~GHz lines are the only two spatially resolved water lines detected in the HL~Tau disk to date. However, water has been detected in this disk before at lower angular resolution with other observatories such as \textit{Herschel}, Gemini North, the Submillimeter Array (SMA), and ALMA \citep{RiviereMarichalar2012, Kristensen2016, AlonsoMartinez2017, Salyk2019, Facchini2024}. A rotational diagram of these lines assuming an ortho-to-para ratio of 3 is presented in the left panel of Fig.~\ref{fig:rot_dia_unres}. The \ce{H2O} line detected with the SMA is not included in this diagram as it is shifted by -20~km~s$^{-1}$ with respect to the source velocity of HL~Tau whereas this line observed with ALMA is centered around the systemic velocity, suggesting that the former does not trace the same emission as the other water lines \citep{Kristensen2016, Facchini2024}. The column densities of the upper level are derived by assuming a 17~au emitting region in the disk frame, following the upper limit on the snowline location \citep{Facchini2024}. A different emitting region will shift all datapoints by a constant factor that is the ratio of the emitting regions but it does not change the relative differences between the datapoints. An overview of all the lines and the line parameters used is presented in Table~\ref{tab:all_H2O}. 

The rotational diagram shows a 2-3 order of magnitude difference between the column density of the \ce{H2O} lines observed with ALMA (cyan) compared to those observed with \textit{Herschel} (pink) despite the upper energy levels being similar (Fig.~\ref{fig:rot_dia_unres} left). Two possible causes for this difference include masing of the 183, 321, and 325~GHz lines seen with ALMA and water emission being hidden by optically thick dust. These ALMA lines are known to be potential masers if the gas density is lower than $\sim 10^{10}$~cm$^{-3}$ and in case of the 321~GHz line, a temperature exceeding $\sim 500$~K \citep{Gray2016}. In this case, the assumption of LTE to convert the line flux to a column density is no longer valid and the derived column density overestimates the true column density due to the population inversion. 

To investigate the second option, we made a simple thermochemical DALI model of a full disk that reproduces the continuum and CO isotopologue flux seen in the HL~Tau disk with ALMA within a factor of 2 for most disk regions (see App.~\ref{app:DALI} for details). We note that the first dust gap in the HL~Tau disk lies at 13~au which is outside the predicted water snowline location (2~au) for this model (black contour in Fig.~\ref{fig:multiplot_H2O}). For each of the lines in Fig.~\ref{fig:rot_dia_unres}, the position in radius $r$ and height $z$ of the $\tau_{\rm dust}=1$ surface is computed. As the continuum optical depth decreases with increasing wavelength, this surface shifts to layers deeper in the disk, revealing more of the water reservoir at longer wavelengths, see Fig.~\ref{fig:multiplot_H2O}. The fraction of the gas-phase \ce{H2O} molecules above that surface ($\mathcal{N}_{\rm above\ \tau_{\rm dust} = 1}$) compared to the total number of gas-phase \ce{H2O} molecules in the disk ($\mathcal{N}_{\rm tot\ in\ disk}$) is a measure for the amount of water hidden below the optically thick dust (see Table~\ref{tab:all_H2O}). The \ce{H2O} line fluxes themselves are not compared as a number of lines are predicted to be masing in some regions of the disk, which is artificially quenched in DALI by limiting their optical depth. 

At mid-IR wavelengths, the continuum optical depth hides most of the thermally desorbed water reservoir and only 0.02\% of the total gas-phase water reservoir is accessible with observations (Fig.~\ref{fig:multiplot_H2O} left). This is similar to what was found in \citet{Houge2025} who used dust evolution models to investigate the observable water reservoir as function of time. At far-IR wavelengths such as those probed by \textit{Herschel} PACS, the optically thick dust in the model hides the water below $z/r$ of $\sim0.15-0.2$ depending on the wavelength within the far-IR (Fig.~\ref{fig:multiplot_H2O} middle). As the gas density is much higher in the disk midplane than in the surface layers, even in the far-IR, only 0.2-2\% of the water reservoir in the model resides above the optically thick dust. Only in the (sub-)mm, the modelled $\tau_{\rm dust}=1$ surface is located deep inside the disk allowing observations to probe the water snowline position close to the disk midplane (Fig.~\ref{fig:multiplot_H2O} right). 

The right panel in Fig.~\ref{fig:rot_dia_unres} shows the same rotational diagram as in the left panel but with the correction for the fraction of the water reservoir that is visible applied to each line. The \ce{H2O} lines observed with ALMA only move up by a factor of a few as 35-70\% of the gas-phase water reservoir is visible to observations at those frequencies. In contrast, the mid- and far-IR datapoints move up by orders of magnitude. The difference in the $N_u/g_u$ between the (sub-)mm and far-IR lines with similar $E_u$ is much smaller (a factor of a few) after correcting for the optical depth of the continuum. Therefore, the 183, 321, and 325~GHz lines in this work and \citet{Facchini2024} are unlikely to be strongly masing.

\subsection{\ce{H2O} vs \ce{H^13CO+}}

\begin{figure}
\centering
\includegraphics[width=\hsize]{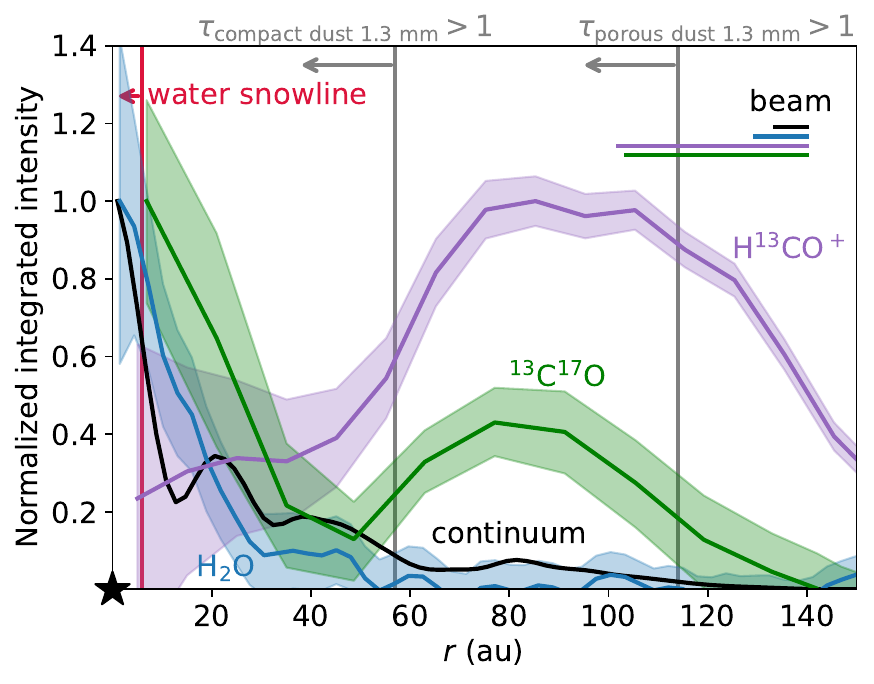}
  \caption{Azimuthally averaged radial profiles of the \ce{H2O} 183~GHz line imaged with \texttt{robust} = 0, the \ce{H^13CO+} $J=2-1$ line, the \ce{^13C^17O} $J=3-2$ line by \citet{Booth2020}, and the 1.7~mm continuum. The grey vertical lines indicate where the 1.3~mm continuum emission becomes optically thick assuming compact grains or grains with a 90\% porosity \citep{GuerraAlvarado2024} and the upper limit on the water snowline location is indicated with the vertical red line.}  
     \label{fig:azi_avg_H13CO+}
\end{figure}

\begin{figure*}
\centering
\includegraphics[width=\hsize]{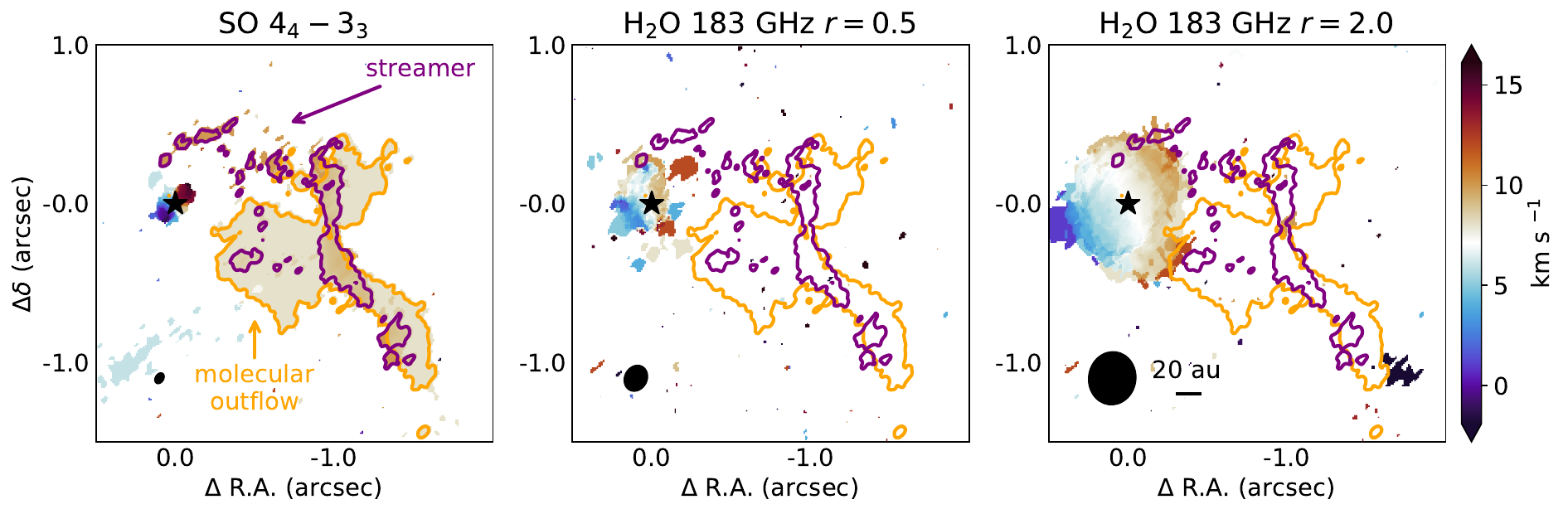}
  \caption{Moment 1 maps of the SO $4_4-3_3$ transition (left) and the \ce{H2O} 183~GHz line imaged with a \texttt{robust} parameter of 0.5 (middle) and 2.0 (right). The orange and purple contours indicate the emission above 5$\sigma$ in the SO channel at 8 and 10~km~s$^{-1}$ tracing primarily the molecular outflow and the streamer, respectively where 1$\sigma$ corresponds to 0.3~mJy~beam$^{-1}$. The beam of the respective moment 1 maps is indicated in the bottom left corner of each panel and a 20~au scale bar is shown in the right panel.}  
     \label{fig:mom1_H2O_SO}
\end{figure*}

The radial profiles of the high resolution \ce{H2O} data suggest that the water snowline is located inside the inner beam of the observations. Another tracer of the water snowline is \ce{H^13CO+}, where ring-shaped emission is expected outside the water snowline as gas-phase water destroys \ce{HCO+} \citep{Leemker2021}. Other possible causes for a central hole in the \ce{H^13CO+} emission include excitation, optically thick dust hiding the line emission, and absorption by the surrounding envelope. In this subsection, we present the \ce{H^13CO+} $J=2-1$ line covered by the long baseline ALMA data (Table~\ref{tab:det_ALMA}). The integrated intensity map, the moment~1 map, and the spectrum are presented in Fig.~\ref{fig:mom01_H13CO+} and \ref{fig:spec_H13CO+} in App.~\ref{app:maps_and_spectra}. 

The normalized azimuthally averaged radial profile of the \ce{H^13CO+} emission in Fig.~\ref{fig:azi_avg_H13CO+} shows that the emission is indeed ring-shaped and peaks at $\sim80$~au from the central star. Even though HL~Tau is embedded in an envelope, no strong absorption feature is seen at the systemic velocity in the \ce{H2O} and \ce{H^13CO+} spectra (see Fig.~\ref{fig:spec_H2O} and the right panel Fig.~\ref{fig:spec_H13CO+}). Therefore, the effect of the envelope on the central hole in \ce{H^13CO+} emission is at most marginal. 

To investigate the effect of optically thick dust hiding the \ce{H^13CO+} emission in the center, the \ce{H^13CO+} $J=2-1$ (173.507~GHz) radial profile is compared to that of the \ce{^13C^17O} $J=3-2$ emission at 321.852~GHz presented in \citet{Booth2020}. Under the assumption that the underlying \ce{^13C^17O} column density profile is independent of radius and the line is optically thin, the substructures seen in the \ce{^13C^17O} emission trace the effects of the optically thick dust.

Interestingly, the \ce{H^13CO+} and \ce{^13C^17O} emission both exhibit a ring at roughly the same radius of $\sim$80~au, similar to the ring detected in \ce{H2CO}, \ce{H2CS}, and CS \citep{Garufi2021}. Inside this radius, both profiles decline, with \ce{H^13CO+} continuing to decrease steadily toward the central star, while the \ce{^13C^17O} emission rises inward of the 40~au gap. Dust modeling of the HL Tau disk indicates that the continuum becomes optically thick inside $\sim$60~au for compact grains, or inside $\sim$115 au if the grains are $\sim$90\% porous \citep{CarrascoGonzalez2019, GuerraAlvarado2024, Ueda2025}. As \ce{H^13CO+}, which is likely optically thin or close to optically thin, primarily emits from the dense disk midplane, the optically thick dust absorbs these line photons, producing the observed central hole \citep[e.g.,][]{Bosman2021_MAPS15}.

The contrasting morphology (\ce{H^13CO+} showing a central cavity, and \ce{H2O} and \ce{^13C^17O} exhibiting centrally peaked emission) can be explained by the disk's three-dimensional geometry and the interaction between line and continuum opacity. Line emission from higher disk layers can still escape if the molecule remains abundant above the dust $\tau = 1$ surface. 
\citet{Isella2016} and \citet{Weaver2018} showed that for an optically thick line, the intensity after continuum subtraction  reflects the difference between gas and dust temperatures as the gas absorbs part of the continuum emission.
Because the gas in the layers where \ce{H2O} and \ce{^13C^17O} originate is warmer than the dust in the cooler midplane, line photons emitted from above the optically thick dust can still be observed as a central peak in azimuthally averaged radial profiles. The absence of a similar increase in the \ce{H^13CO+} profile indicates that this line does not become optically thick above the $\tau_{\rm dust} = 1$ surface, despite model predictions that the \ce{HCO+} column density increases to several $\times 10^{13}$~cm$^{-2}$ outside the water snowline \citep{Leemker2021} and despite its lower frequency compared to the \ce{^13C^17O} transition.

In summary, the disk's 3D structure hides emission from molecules located below the $\tau_{\rm dust} = 1$ surface (e.g., \ce{H^13CO+}) while still allowing photons from species abundant above this layer such as \ce{H2O} and \ce{^13C^17O} to escape. Therefore, the central depression in \ce{H^13CO+} is likely caused by optically thick dust rather than by the water snowline.

\section{Discussion} \label{sec:disc}

\subsection{Location of the water snowline in the HL tau disk}

The midplane dust temperature derived from modelling high-resolution, multi-wavelength continuum observations shows that the water snowline is located around 5~au in the disk midplane \citep{CarrascoGonzalez2019, GuerraAlvarado2024, Ueda2025}. The similarity in the location of the midplane water snowline derived from the dust and the upper limit of $\sim 6$~au based on the \ce{H2O} emission suggests that the water snowline is close to vertical up to the height where the \ce{H2O} 183~GHz line becomes optically thick as the ALMA observations likely do not trace down to the disk midplane. 

The diffuse \ce{H2O} emission seen outside the central beam likely traces hot water at an excitation temperature of $\sim$350~K, consistent with that expected for the reformed water reservoir in the disk surface ($T>300$~K). The presence of this reservoir is further supported by the highly excited \ce{H2O} lines observed with Gemini North ($E_u =$ 4945-5781~K). The line profiles of these transitions are consistent with emission up to at least 5~au where the drop off in the radial profile might be due to the water snowline \citep{Salyk2019}. Finally, the slight asymmetry towards the north/ north east seen in the \ce{H2O} 183~GHz line emission at moderate spatial resolution could be due to projection effects of this elevated layer, a disk wind traced by \ce{H2O} emission, or due to the infalling streamer impacting the HL~Tau disk in the north.

\subsection{\ce{H2O} and SO}

The HL~Tau disk is impacted in the north by an infalling streamer seen in \ce{HCO+} \citep{Yen2019, Gupta2024}. The resulting shock releases energy into the disk, liberating SO and \ce{SO2} \citep{Garufi2021}.
The lower limit of the excitation temperature of the \ce{SO2} in the HL~Tau inner disk of $>350$~K is consistent with the excitation temperature of the \ce{H2O} obtained from the radially resolved rotational diagram, suggesting a possible link between the water and the shock tracers SO and \ce{SO2}.

In Fig.~\ref{fig:mom1_H2O_SO}, we present a comparison of the velocity structure of the cold SO $4_4-3_3$ transition ($E_u = $~33.7~K) covered by the long baseline data and the \ce{H2O} 183~GHz line. The SO shows compact emission on the scale of the disk ($r \lesssim 0\farcs17$) and extended emission to the south west out to $\sim2$". The latter consists of two components: a streamer and a molecular outflow or wind (see also Fig.~\ref{fig:mom0_SO}). The streamer is indicated with the purple contour and overlaps with that seen in \ce{HCO+}. The outflow or wind component is indicated with the orange contour and is similar to that seen in CO \citep{Lumbreras2014, Bacciotti2025}. 

The velocity structure of the SO emission is compared to that of the water imaged with a \texttt{robust} parameter of 0.5 and 2.0 in the middle and right panels of Fig.~\ref{fig:mom1_H2O_SO}. The high resolution \ce{H2O} map in the middle panel shows complex structures in the velocity. A velocity gradient is seen along the disk major axis indicating Keplerian rotation, similar to the SO emission in the inner $0\farcs17$. In the north, the \ce{H2O} moment~1 map imaged with a small \texttt{robust} parameter of 0.5 shows some red-shifted emission aligning with the red-shifted emission seen in the SO streamer at that location. This hints that some of the \ce{H2O} emission might be due to the shock impacting the disk. In addition, some red-shifted \ce{H2O} emission is seen in the south when imaged at moderate spatial resolution (\texttt{robust} = 2.0). As this image probes the diffuse water in the disk surface layers, the red-shifted emission could indicate a connection to the outflow or a disk wind \citep{Lumbreras2014, Bacciotti2025}. All in all, the water shows a highly complex velocity structure with indications of Keplerian rotation in the center, and a possible connection to a streamer, molecular outflow or wind.

\section{Conclusions} \label{sec:concl}

In this work, we present the first high $\sim0\farcs08$ resolution images of the \ce{H2O} line at 183~GHz in the HL~Tau disk. We derived the radially resolved temperature profile of the water emission through a rotational diagram analysis with the 321~GHz line of \ce{H2O}. In addition, we quantified the fraction of the water reservoir visible at sub-mm wavelengths and compared this to that at far- and mid-infrared wavelengths. Finally, \ce{H^13CO+} emission was examined as a snowline tracer in this disk. In summary, we find that:

\begin{itemize}
    \item The \ce{H2O} 183~GHz line shows a central, compact and optically thick component and diffuse emission that extends out to $\sim 75$~au. The emission in the central beam probes the water inside the water snowline at $\lesssim 6$~au at the height probed by the observations. The disk midplane is hidden due to optical depth effects of the water and dust.  
    \item A disk-integrated rotational diagram comparing \ce{H2O} lines seen at (sub-)mm wavelengths to those at far- and mid-IR wavelengths demonstrates that the lines seen with ALMA probe a 2-3 orders of magnitude larger column density of water. This is primarily due to optically thick dust hiding $98-99.98$\% of the water reservoir at far-and mid-IR wavelengths.
    \item Correcting the disk-integrated rotational diagram for optically thick dust shows that the \ce{H2O} 183, 321, and 325~GHz lines are not strongly masing. 
    \item Observing the \ce{H2O} 183~GHz line directly is a better probe of the water snowline than the \ce{H^13CO+} $J=2-1$ line due to optically thick dust. 
    \item Some of the \ce{H2O} emission might originate due to the infalling streamer impacting the disk which for the first time detected in SO. This cold SO $4_4-3_3$ transition traces the disk, the infalling streamer seen in \ce{HCO+}, and the molecular outflow \citep{Lumbreras2014, Yen2019, Bacciotti2025}. 
\end{itemize}

This work demonstrates that observing the \ce{H2O} 183~GHz line at high spatial resolution is the most effective strategy to directly trace the water snowline in disks. Other tracers such as \ce{H^13CO+} and \ce{CH3OH} emission are affected by optically thick dust and/ or not detected in this disk \citep{Soaveinprep}. Similarly, water lines in the mid- and far-IR regime trace the water snowline in the disk surface layers due to the high continuum optical depth. Instead, the 183~GHz line probes much closer to the disk midplane, providing the best strategy to derive the water snowline.

\begin{acknowledgements}
We thank the referee Shigehisa Takakuwa for the constructive comments. 
M.L. acknowledges assistance from Allegro, the European ALMA Regional Center node in the Netherlands. M.L., S.F., and L.R. are funded by the European Union (ERC, UNVEIL, 101076613). Views and opinions expressed are however those of the author(s) only and do not necessarily reflect those of the European Union or the European Research Council. Neither the European Union nor the granting authority can be held responsible for them. S.F. acknowledges financial contribution from PRIN-MUR 2022YP5ACE. 
P.C. acknowledges support by the ANID BASAL project FB210003.
M.B. has received funding from the European Research Council (ERC) under the European Union’s Horizon 2020 research and innovation programme (PROTOPLANETS, grant agreement No. 101002188).
This paper makes use of the following ALMA data: ADS/JAO.ALMA\#2017.1.01178.S, \#2017.1.01562.S, \#2021.1.01310.S, and \#2022.1.00905.S. ALMA is a partnership of ESO (representing its member states), NSF (USA) and NINS (Japan), together with NRC (Canada), NSTC and ASIAA (Taiwan), and KASI (Republic of Korea), in cooperation with the Republic of Chile. The Joint ALMA Observatory is operated by ESO, AUI/NRAO and NAOJ. 
\end{acknowledgements}

%
%

\bibliographystyle{aa} 
\bibliography{refs.bib} 

@ARTICLE{ALMA2015,
       author = {{ALMA Partnership} and {Brogan}, C.~L. and {P{\'e}rez}, L.~M. and {Hunter}, T.~R. and {Dent}, W.~R.~F. and {Hales}, A.~S. and {Hills}, R.~E. and {Corder}, S. and {Fomalont}, E.~B. and {Vlahakis}, C. and {Asaki}, Y. and {Barkats}, D. and {Hirota}, A. and {Hodge}, J.~A. and {Impellizzeri}, C.~M.~V. and {Kneissl}, R. and {Liuzzo}, E. and {Lucas}, R. and {Marcelino}, N. and {Matsushita}, S. and {Nakanishi}, K. and {Phillips}, N. and {Richards}, A.~M.~S. and {Toledo}, I. and {Aladro}, R. and {Broguiere}, D. and {Cortes}, J.~R. and {Cortes}, P.~C. and {Espada}, D. and {Galarza}, F. and {Garcia-Appadoo}, D. and {Guzman-Ramirez}, L. and {Humphreys}, E.~M. and {Jung}, T. and {Kameno}, S. and {Laing}, R.~A. and {Leon}, S. and {Marconi}, G. and {Mignano}, A. and {Nikolic}, B. and {Nyman}, L. -A. and {Radiszcz}, M. and {Remijan}, A. and {Rod{\'o}n}, J.~A. and {Sawada}, T. and {Takahashi}, S. and {Tilanus}, R.~P.~J. and {Vila Vilaro}, B. and {Watson}, L.~C. and {Wiklind}, T. and {Akiyama}, E. and {Chapillon}, E. and {de Gregorio-Monsalvo}, I. and {Di Francesco}, J. and {Gueth}, F. and {Kawamura}, A. and {Lee}, C. -F. and {Nguyen Luong}, Q. and {Mangum}, J. and {Pietu}, V. and {Sanhueza}, P. and {Saigo}, K. and {Takakuwa}, S. and {Ubach}, C. and {van Kempen}, T. and {Wootten}, A. and {Castro-Carrizo}, A. and {Francke}, H. and {Gallardo}, J. and {Garcia}, J. and {Gonzalez}, S. and {Hill}, T. and {Kaminski}, T. and {Kurono}, Y. and {Liu}, H. -Y. and {Lopez}, C. and {Morales}, F. and {Plarre}, K. and {Schieven}, G. and {Testi}, L. and {Videla}, L. and {Villard}, E. and {Andreani}, P. and {Hibbard}, J.~E. and {Tatematsu}, K.},
        title = "{The 2014 ALMA Long Baseline Campaign: First Results from High Angular Resolution Observations toward the HL Tau Region}",
      journal = {\apjl},
         year = 2015,
        month = jul,
       volume = {808},
       number = {1},
          eid = {L3},
        pages = {L3},
          doi = {10.1088/2041-8205/808/1/L3},
archivePrefix = {arXiv},
       eprint = {1503.02649},
 primaryClass = {astro-ph.SR},
       adsurl = {https://ui.adsabs.harvard.edu/abs/2015ApJ...808L...3A}
}

@ARTICLE{AlonsoMartinez2017,
       author = {{Alonso-Mart{\'\i}nez}, M. and {Riviere-Marichalar}, P. and {Meeus}, G. and {Kamp}, I. and {Fang}, M. and {Podio}, L. and {Dent}, W.~R.~F. and {Eiroa}, C.},
        title = "{Herschel GASPS spectral observations of T Tauri stars in Taurus. Unraveling far-infrared line emission from jets and discs}",
      journal = {\aap},
         year = 2017,
        month = jul,
       volume = {603},
          eid = {A138},
        pages = {A138},
          doi = {10.1051/0004-6361/201629005},
archivePrefix = {arXiv},
       eprint = {1704.04834},
 primaryClass = {astro-ph.SR},
       adsurl = {https://ui.adsabs.harvard.edu/abs/2017A&A...603A.138A}
}

@ARTICLE{Andrews2018,
       author = {{Andrews}, Sean M. and {Huang}, Jane and {P{\'e}rez}, Laura M. and {Isella}, Andrea and {Dullemond}, Cornelis P. and {Kurtovic}, Nicol{\'a}s T. and {Guzm{\'a}n}, Viviana V. and {Carpenter}, John M. and {Wilner}, David J. and {Zhang}, Shangjia and {Zhu}, Zhaohuan and {Birnstiel}, Tilman and {Bai}, Xue-Ning and {Benisty}, Myriam and {Hughes}, A. Meredith and {{\"O}berg}, Karin I. and {Ricci}, Luca},
        title = "{The Disk Substructures at High Angular Resolution Project (DSHARP). I. Motivation, Sample, Calibration, and Overview}",
      journal = {\apjl},
         year = 2018,
        month = dec,
       volume = {869},
       number = {2},
          eid = {L41},
        pages = {L41},
          doi = {10.3847/2041-8213/aaf741},
archivePrefix = {arXiv},
       eprint = {1812.04040},
 primaryClass = {astro-ph.SR},
       adsurl = {https://ui.adsabs.harvard.edu/abs/2018ApJ...869L..41A}
}

@ARTICLE{Bacciotti2025,
       author = {{Bacciotti}, F. and {Nony}, T. and {Podio}, L. and {Dougados}, C. and {Garufi}, A. and {Cabrit}, S. and {Codella}, C. and {Zimniak}, N. and {Ferreira}, J.},
        title = "{ALMA chemical survey of disk-outflow sources in Taurus (ALMA-DOT) VII: the layered molecular outflow from HL Tau and its relationship with the ringed disk}",
      journal = {arXiv e-prints},
         year = 2025,
        month = jan,
          eid = {arXiv:2501.03920},
        pages = {arXiv:2501.03920},
          doi = {10.48550/arXiv.2501.03920},
archivePrefix = {arXiv},
       eprint = {2501.03920},
 primaryClass = {astro-ph.GA},
       adsurl = {https://ui.adsabs.harvard.edu/abs/2025arXiv250103920B}
}

@ARTICLE{Banzatti2023,
       author = {{Banzatti}, Andrea and {Pontoppidan}, Klaus M. and {Carr}, John S. and {Jellison}, Evan and {Pascucci}, Ilaria and {Najita}, Joan R. and {Romero-Mirza}, Carlos E. and {{\"O}berg}, Karin I. and {Kalyaan}, Anusha and {Pinilla}, Paola and {Krijt}, Sebastiaan and {Long}, Feng and {Lambrechts}, Michiel and {Rosotti}, Giovanni and {Herczeg}, Gregory J. and {Salyk}, Colette and {Zhang}, Ke and {Bergin}, Edwin A. and {Ballering}, Nicholas P. and {Meyer}, Michael R. and {Bruderer}, Simon and {Jdiscs Collaboration}},
        title = "{JWST Reveals Excess Cool Water near the Snow Line in Compact Disks, Consistent with Pebble Drift}",
      journal = {\apjl},
         year = 2023,
        month = nov,
       volume = {957},
       number = {2},
          eid = {L22},
        pages = {L22},
          doi = {10.3847/2041-8213/acf5ec},
archivePrefix = {arXiv},
       eprint = {2307.03846},
 primaryClass = {astro-ph.EP},
       adsurl = {https://ui.adsabs.harvard.edu/abs/2023ApJ...957L..22B}
}

@ARTICLE{Banzatti2025,
       author = {{Banzatti}, Andrea and {Salyk}, Colette and {Pontoppidan}, Klaus M. and {Carr}, John S. and {Zhang}, Ke and {Arulanantham}, Nicole and {Krijt}, Sebastiaan and {{\"O}berg}, Karin I. and {Cleeves}, L. Ilsedore and {Najita}, Joan R. and {Pascucci}, Ilaria and {Blake}, Geoffrey A. and {Romero-Mirza}, Carlos E. and {Bergin}, Edwin A. and {Cieza}, Lucas A. and {Pinilla}, Paola and {Long}, Feng and {Mallaney}, Patrick and {Xie}, Chengyan and {Waggoner}, Abygail R. and {Kaeufer}, Till and {The Jdiscs Collaboration}},
        title = "{Water in Protoplanetary Disks with JWST-MIRI: Spectral Excitation Atlas and Radial Distribution from Temperature Diagnostic Diagrams and Doppler Mapping}",
      journal = {\aj},
         year = 2025,
        month = mar,
       volume = {169},
       number = {3},
          eid = {165},
        pages = {165},
          doi = {10.3847/1538-3881/ada962},
archivePrefix = {arXiv},
       eprint = {2409.16255},
 primaryClass = {astro-ph.EP},
       adsurl = {https://ui.adsabs.harvard.edu/abs/2025AJ....169..165B}
}

@ARTICLE{Barber2006,
       author = {{Barber}, R.~J. and {Tennyson}, J. and {Harris}, G.~J. and {Tolchenov}, R.~N.},
        title = "{A high-accuracy computed water line list}",
      journal = {\mnras},
         year = 2006,
        month = may,
       volume = {368},
       number = {3},
        pages = {1087-1094},
          doi = {10.1111/j.1365-2966.2006.10184.x},
archivePrefix = {arXiv},
       eprint = {astro-ph/0601236},
 primaryClass = {astro-ph},
       adsurl = {https://ui.adsabs.harvard.edu/abs/2006MNRAS.368.1087B}
}

@ARTICLE{Beck2010,
       author = {{Beck}, Tracy L. and {Bary}, Jeffery S. and {McGregor}, Peter J.},
        title = "{Spatially Extended Brackett Gamma Emission in the Environments of Young Stars}",
      journal = {\apj},
         year = 2010,
        month = oct,
       volume = {722},
       number = {2},
        pages = {1360-1372},
          doi = {10.1088/0004-637X/722/2/1360},
archivePrefix = {arXiv},
       eprint = {1008.4101},
 primaryClass = {astro-ph.SR},
       adsurl = {https://ui.adsabs.harvard.edu/abs/2010ApJ...722.1360B}
}

@ARTICLE{Bergin1998,
       author = {{Bergin}, Edwin A. and {Melnick}, Gary J. and {Neufeld}, David A.},
        title = "{The Postshock Chemical Lifetimes of Outflow Tracers and a Possible New Mechanism to Produce Water Ice Mantles}",
      journal = {\apj},
         year = 1998,
        month = may,
       volume = {499},
       number = {2},
        pages = {777-792},
          doi = {10.1086/305656},
archivePrefix = {arXiv},
       eprint = {astro-ph/9803330},
 primaryClass = {astro-ph},
       adsurl = {https://ui.adsabs.harvard.edu/abs/1998ApJ...499..777B}
}

@ARTICLE{Booth2020,
       author = {{Booth}, Alice S. and {Ilee}, John D.},
        title = "{$^{13}$C$^{17}$O suggests gravitational instability in the HL Tau disc}",
      journal = {\mnras},
         year = 2020,
        month = mar,
       volume = {493},
       number = {1},
        pages = {L108-L113},
          doi = {10.1093/mnrasl/slaa014},
archivePrefix = {arXiv},
       eprint = {2001.07550},
 primaryClass = {astro-ph.SR},
       adsurl = {https://ui.adsabs.harvard.edu/abs/2020MNRAS.493L.108B}
}

@ARTICLE{Bosman2021,
       author = {{Bosman}, Arthur D. and {Bergin}, Edwin A.},
        title = "{Reimagining the Water Snowline}",
      journal = {\apjl},
         year = 2021,
        month = sep,
       volume = {918},
       number = {1},
          eid = {L10},
        pages = {L10},
          doi = {10.3847/2041-8213/ac1db1},
archivePrefix = {arXiv},
       eprint = {2108.07303},
 primaryClass = {astro-ph.EP},
       adsurl = {https://ui.adsabs.harvard.edu/abs/2021ApJ...918L..10B}
}

@ARTICLE{Bosman2021_MAPS15,
       author = {{Bosman}, Arthur D. and {Bergin}, Edwin A. and {Loomis}, Ryan A. and {Andrews}, Sean M. and {van't Hoff}, Merel L.~R. and {Teague}, Richard and {{\"O}berg}, Karin I. and {Guzm{\'a}n}, Viviana V. and {Walsh}, Catherine and {Aikawa}, Yuri and {Alarc{\'o}n}, Felipe and {Bae}, Jaehan and {Bergner}, Jennifer B. and {Booth}, Alice S. and {Cataldi}, Gianni and {Cleeves}, L. Ilsedore and {Czekala}, Ian and {Huang}, Jane and {Ilee}, John D. and {Law}, Charles J. and {Le Gal}, Romane and {Liu}, Yao and {Long}, Feng and {M{\'e}nard}, Fran{\c{c}}ois and {Nomura}, Hideko and {P{\'e}rez}, Laura M. and {Qi}, Chunhua and {Schwarz}, Kamber R. and {Sierra}, Anibal and {Tsukagoshi}, Takashi and {Yamato}, Yoshihide and {Wilner}, David J. and {Zhang}, Ke},
        title = "{Molecules with ALMA at Planet-forming Scales (MAPS). XV. Tracing Protoplanetary Disk Structure within 20 au}",
      journal = {\apjs},
         year = 2021,
        month = nov,
       volume = {257},
       number = {1},
          eid = {15},
        pages = {15},
          doi = {10.3847/1538-4365/ac1433},
archivePrefix = {arXiv},
       eprint = {2109.06223},
 primaryClass = {astro-ph.EP},
       adsurl = {https://ui.adsabs.harvard.edu/abs/2021ApJS..257...15B}
}

@ARTICLE{Bosman2022,
       author = {{Bosman}, Arthur D. and {Bergin}, Edwin A. and {Calahan}, Jenny and {Duval}, Sara E.},
        title = "{Water UV-shielding in the Terrestrial Planet-forming Zone: Implications from Water Emission}",
      journal = {\apjl},
         year = 2022,
        month = may,
       volume = {930},
       number = {2},
          eid = {L26},
        pages = {L26},
          doi = {10.3847/2041-8213/ac66ce},
archivePrefix = {arXiv},
       eprint = {2204.07108},
 primaryClass = {astro-ph.EP},
       adsurl = {https://ui.adsabs.harvard.edu/abs/2022ApJ...930L..26B}
}

@ARTICLE{Briceno2002,
       author = {{Brice{\~n}o}, C{\'e}sar and {Luhman}, K.~L. and {Hartmann}, Lee and {Stauffer}, John R. and {Kirkpatrick}, J. Davy},
        title = "{The Initial Mass Function in the Taurus Star-forming Region}",
      journal = {\apj},
         year = 2002,
        month = nov,
       volume = {580},
       number = {1},
        pages = {317-335},
          doi = {10.1086/343127},
       adsurl = {https://ui.adsabs.harvard.edu/abs/2002ApJ...580..317B}
}

@ARTICLE{Bruderer2009,
       author = {{Bruderer}, S. and {Doty}, S.~D. and {Benz}, A.~O.},
        title = "{Chemical Modeling of Young Stellar Objects. I. Method and Benchmarks}",
      journal = {\apjs},
         year = 2009,
        month = aug,
       volume = {183},
       number = {2},
        pages = {179-196},
          doi = {10.1088/0067-0049/183/2/179},
archivePrefix = {arXiv},
       eprint = {0906.0584},
 primaryClass = {astro-ph.SR},
       adsurl = {https://ui.adsabs.harvard.edu/abs/2009ApJS..183..179B}
}

@ARTICLE{Bruderer2012,
       author = {{Bruderer}, S. and {van Dishoeck}, E.~F. and {Doty}, S.~D. and {Herczeg}, G.~J.},
        title = "{The warm gas atmosphere of the HD 100546 disk seen by Herschel. Evidence of a gas-rich, carbon-poor atmosphere?}",
      journal = {\aap},
         year = 2012,
        month = may,
       volume = {541},
          eid = {A91},
        pages = {A91},
          doi = {10.1051/0004-6361/201118218},
archivePrefix = {arXiv},
       eprint = {1201.4860},
 primaryClass = {astro-ph.SR},
       adsurl = {https://ui.adsabs.harvard.edu/abs/2012A&A...541A..91B}
}

@ARTICLE{Bruderer2013,
       author = {{Bruderer}, Simon},
        title = "{Survival of molecular gas in cavities of transition disks. I. CO}",
      journal = {\aap},
         year = 2013,
        month = nov,
       volume = {559},
          eid = {A46},
        pages = {A46},
          doi = {10.1051/0004-6361/201321171},
archivePrefix = {arXiv},
       eprint = {1308.2966},
 primaryClass = {astro-ph.SR},
       adsurl = {https://ui.adsabs.harvard.edu/abs/2013A&A...559A..46B}
}

@ARTICLE{Carr2018,
       author = {{Carr}, John S. and {Najita}, Joan R. and {Salyk}, Colette},
        title = "{Measuring the Water Snow Line in a Protoplanetary Disk}",
      journal = {Research Notes of the American Astronomical Society},
         year = 2018,
        month = sep,
       volume = {2},
       number = {3},
          eid = {169},
        pages = {169},
          doi = {10.3847/2515-5172/aadfe7},
       adsurl = {https://ui.adsabs.harvard.edu/abs/2018RNAAS...2..169C}
}

@ARTICLE{CarrascoGonzalez2019,
       author = {{Carrasco-Gonz{\'a}lez}, Carlos and {Sierra}, Anibal and {Flock}, Mario and {Zhu}, Zhaohuan and {Henning}, Thomas and {Chandler}, Claire and {Galv{\'a}n-Madrid}, Roberto and {Mac{\'\i}as}, Enrique and {Anglada}, Guillem and {Linz}, Hendrik and {Osorio}, Mayra and {Rodr{\'\i}guez}, Luis F. and {Testi}, Leonardo and {Torrelles}, Jos{\'e} M. and {P{\'e}rez}, Laura and {Liu}, Yao},
        title = "{The Radial Distribution of Dust Particles in the HL Tau Disk from ALMA and VLA Observations}",
      journal = {\apj},
         year = 2019,
        month = sep,
       volume = {883},
       number = {1},
          eid = {71},
        pages = {71},
          doi = {10.3847/1538-4357/ab3d33},
archivePrefix = {arXiv},
       eprint = {1908.07140},
 primaryClass = {astro-ph.EP},
       adsurl = {https://ui.adsabs.harvard.edu/abs/2019ApJ...883...71C}
}

@software{carta,
  author       = {Angus Comrie and
                  Kuo-Song Wang and
                  Yu-Hsuan Hwang and
                  Anthony Moraghan and
                  Pamela Harris and
                  Adrianna Pińska and
                  Carli Raul-Omar and
                  Cheng-Chin Chiang and
                  Ming-Yi Lin and
                  Tien-Hao Chang and
                  Rob Simmonds},
  title        = {CARTA: The Cube Analysis and Rendering Tool for
                   Astronomy
                  },
  month        = jan,
  year         = 2024,
  publisher    = {Zenodo},
  version      = {4.1.0},
  doi          = {10.5281/zenodo.15172686},
  url          = {https://doi.org/10.5281/zenodo.15172686},
}

@ARTICLE{casa2022,
       author = {{CASA Team} and {Bean}, Ben and {Bhatnagar}, Sanjay and {Castro}, Sandra and {Donovan Meyer}, Jennifer and {Emonts}, Bjorn and {Garcia}, Enrique and {Garwood}, Robert and {Golap}, Kumar and {Gonzalez Villalba}, Justo and {Harris}, Pamela and {Hayashi}, Yohei and {Hoskins}, Josh and {Hsieh}, Mingyu and {Jagannathan}, Preshanth and {Kawasaki}, Wataru and {Keimpema}, Aard and {Kettenis}, Mark and {Lopez}, Jorge and {Marvil}, Joshua and {Masters}, Joseph and {McNichols}, Andrew and {Mehringer}, David and {Miel}, Renaud and {Moellenbrock}, George and {Montesino}, Federico and {Nakazato}, Takeshi and {Ott}, Juergen and {Petry}, Dirk and {Pokorny}, Martin and {Raba}, Ryan and {Rau}, Urvashi and {Schiebel}, Darrell and {Schweighart}, Neal and {Sekhar}, Srikrishna and {Shimada}, Kazuhiko and {Small}, Des and {Steeb}, Jan-Willem and {Sugimoto}, Kanako and {Suoranta}, Ville and {Tsutsumi}, Takahiro and {van Bemmel}, Ilse M. and {Verkouter}, Marjolein and {Wells}, Akeem and {Xiong}, Wei and {Szomoru}, Arpad and {Griffith}, Morgan and {Glendenning}, Brian and {Kern}, Jeff},
        title = "{CASA, the Common Astronomy Software Applications for Radio Astronomy}",
      journal = {\pasp},
         year = 2022,
        month = nov,
       volume = {134},
       number = {1041},
          eid = {114501},
        pages = {114501},
          doi = {10.1088/1538-3873/ac9642},
archivePrefix = {arXiv},
       eprint = {2210.02276},
 primaryClass = {astro-ph.IM},
       adsurl = {https://ui.adsabs.harvard.edu/abs/2022PASP..134k4501C}
}

@ARTICLE{Casassus2022,
       author = {{Casassus}, Simon and {C{\'a}rcamo}, Miguel},
        title = "{Variable structure in the PDS 70 disc and uncertainties in radio-interferometric image restoration}",
      journal = {\mnras},
         year = 2022,
        month = jul,
       volume = {513},
       number = {4},
        pages = {5790-5798},
          doi = {10.1093/mnras/stac1285},
archivePrefix = {arXiv},
       eprint = {2204.08589},
 primaryClass = {astro-ph.EP},
       adsurl = {https://ui.adsabs.harvard.edu/abs/2022MNRAS.513.5790C}
}

@ARTICLE{Cheng2022,
       author = {{Cheng}, Y. -C. and {Bockel{\'e}e-Morvan}, D. and {Roos-Serote}, M. and {Crovisier}, J. and {Debout}, V. and {Erard}, S. and {Drossart}, P. and {Leyrat}, C. and {Capaccioni}, F. and {Filacchione}, G. and {Dubernet}, M. -L. and {Encrenaz}, T.},
        title = "{Water ortho-to-para ratio in the coma of comet 67P/Churyumov-Gerasimenko}",
      journal = {\aap},
         year = 2022,
        month = jul,
       volume = {663},
          eid = {A43},
        pages = {A43},
          doi = {10.1051/0004-6361/202142494},
archivePrefix = {arXiv},
       eprint = {2204.13960},
 primaryClass = {astro-ph.EP},
       adsurl = {https://ui.adsabs.harvard.edu/abs/2022A&A...663A..43C}
}

@ARTICLE{Czekala2021,
       author = {{Czekala}, Ian and {Loomis}, Ryan A. and {Teague}, Richard and {Booth}, Alice S. and {Huang}, Jane and {Cataldi}, Gianni and {Ilee}, John D. and {Law}, Charles J. and {Walsh}, Catherine and {Bosman}, Arthur D. and {Guzm{\'a}n}, Viviana V. and {Le Gal}, Romane and {{\"O}berg}, Karin I. and {Yamato}, Yoshihide and {Aikawa}, Yuri and {Andrews}, Sean M. and {Bae}, Jaehan and {Bergin}, Edwin A. and {Bergner}, Jennifer B. and {Cleeves}, L. Ilsedore and {Kurtovic}, Nicolas T. and {M{\'e}nard}, Fran{\c{c}}ois and {Nomura}, Hideko and {P{\'e}rez}, Laura M. and {Qi}, Chunhua and {Schwarz}, Kamber R. and {Tsukagoshi}, Takashi and {Waggoner}, Abygail R. and {Wilner}, David J. and {Zhang}, Ke},
        title = "{Molecules with ALMA at Planet-forming Scales (MAPS). II. CLEAN Strategies for Synthesizing Images of Molecular Line Emission in Protoplanetary Disks}",
      journal = {\apjs},
         year = 2021,
        month = nov,
       volume = {257},
       number = {1},
          eid = {2},
        pages = {2},
          doi = {10.3847/1538-4365/ac1430},
archivePrefix = {arXiv},
       eprint = {2109.06188},
 primaryClass = {astro-ph.EP},
       adsurl = {https://ui.adsabs.harvard.edu/abs/2021ApJS..257....2C}
}

@ARTICLE{Drazkowska2017,
       author = {{Dr{\k{a}}{\.z}kowska}, J. and {Alibert}, Y.},
        title = "{Planetesimal formation starts at the snow line}",
      journal = {\aap},
         year = 2017,
        month = dec,
       volume = {608},
          eid = {A92},
        pages = {A92},
          doi = {10.1051/0004-6361/201731491},
archivePrefix = {arXiv},
       eprint = {1710.00009},
 primaryClass = {astro-ph.EP},
       adsurl = {https://ui.adsabs.harvard.edu/abs/2017A&A...608A..92D}
}

@ARTICLE{Facchini2024,
       author = {{Facchini}, Stefano and {Testi}, Leonardo and {Humphreys}, Elizabeth and {Vander Donckt}, Mathieu and {Isella}, Andrea and {Wrzosek}, Ramon and {Baudry}, Alain and {Gray}, Malcom D. and {Richards}, Anita M.~S. and {Vlemmings}, Wouter},
        title = "{Resolved ALMA observations of water in the inner astronomical units of the HL Tau disk}",
      journal = {Nature Astronomy},
         year = 2024,
        month = may,
       volume = {8},
        pages = {587-595},
          doi = {10.1038/s41550-024-02207-w},
archivePrefix = {arXiv},
       eprint = {2403.00647},
 primaryClass = {astro-ph.EP},
       adsurl = {https://ui.adsabs.harvard.edu/abs/2024NatAs...8..587F}
}

@ARTICLE{Fasano2025,
       author = {{Fasano}, D. and {Benisty}, M. and {Curone}, P. and {Facchini}, S. and {Zagaria}, F. and {Yoshida}, T.~C. and {Doi}, K. and {Sierra}, A. and {Andrews}, S. and {Bae}, J. and {Isella}, A. and {Kurtovic}, N. and {P{\'e}rez}, L.~M. and {Pinilla}, P. and {Rampinelli}, L. and {Teague}, R.},
        title = "{Inner disc and circumplanetary material in the PDS 70 system: Insights from multi-epoch, multi-frequency ALMA observations}",
      journal = {\aap},
         year = 2025,
        month = jul,
       volume = {699},
          eid = {A373},
        pages = {A373},
          doi = {10.1051/0004-6361/202554959},
archivePrefix = {arXiv},
       eprint = {2506.11709},
 primaryClass = {astro-ph.EP},
       adsurl = {https://ui.adsabs.harvard.edu/abs/2025A&A...699A.373F}
}

@ARTICLE{Faure2008,
       author = {{Faure}, A. and {Josselin}, E.},
        title = "{Collisional excitation of water in warm astrophysical media. I. Rate coefficients for rovibrationally excited states}",
      journal = {\aap},
         year = 2008,
        month = dec,
       volume = {492},
       number = {1},
        pages = {257-264},
          doi = {10.1051/0004-6361:200810717},
       adsurl = {https://ui.adsabs.harvard.edu/abs/2008A&A...492..257F}
}

@ARTICLE{Furlan2008,
       author = {{Furlan}, E. and {McClure}, M. and {Calvet}, N. and {Hartmann}, L. and {D'Alessio}, P. and {Forrest}, W.~J. and {Watson}, D.~M. and {Uchida}, K.~I. and {Sargent}, B. and {Green}, J.~D. and {Herter}, T.~L.},
        title = "{Spitzer IRS Spectra and Envelope Models of Class I Protostars in Taurus}",
      journal = {\apjs},
         year = 2008,
        month = may,
       volume = {176},
       number = {1},
        pages = {184-215},
          doi = {10.1086/527301},
archivePrefix = {arXiv},
       eprint = {0711.4038},
 primaryClass = {astro-ph},
       adsurl = {https://ui.adsabs.harvard.edu/abs/2008ApJS..176..184F}
}

@ARTICLE{Galli2018,
       author = {{Galli}, Phillip A.~B. and {Loinard}, Laurent and {Ortiz-L{\'e}on}, Gisela N. and {Kounkel}, Marina and {Dzib}, Sergio A. and {Mioduszewski}, Amy J. and {Rodr{\'\i}guez}, Luis F. and {Hartmann}, Lee and {Teixeira}, Ramachrisna and {Torres}, Rosa M. and {Rivera}, Juana L. and {Boden}, Andrew F. and {Evans}, II, Neal J. and {Brice{\~n}o}, Cesar and {Tobin}, John J. and {Heyer}, Mark},
        title = "{The Gould's Belt Distances Survey (GOBELINS). IV. Distance, Depth, and Kinematics of the Taurus Star-forming Region}",
      journal = {\apj},
         year = 2018,
        month = may,
       volume = {859},
       number = {1},
          eid = {33},
        pages = {33},
          doi = {10.3847/1538-4357/aabf91},
archivePrefix = {arXiv},
       eprint = {1805.09357},
 primaryClass = {astro-ph.SR},
       adsurl = {https://ui.adsabs.harvard.edu/abs/2018ApJ...859...33G}
}

@ARTICLE{Garufi2021,
       author = {{Garufi}, A. and {Podio}, L. and {Codella}, C. and {Fedele}, D. and {Bianchi}, E. and {Favre}, C. and {Bacciotti}, F. and {Ceccarelli}, C. and {Mercimek}, S. and {Rygl}, K. and {Teague}, R. and {Testi}, L.},
        title = "{ALMA chemical survey of disk-outflow sources in Taurus (ALMA-DOT). V. Sample, overview, and demography of disk molecular emission}",
      journal = {\aap},
         year = 2021,
        month = jan,
       volume = {645},
          eid = {A145},
        pages = {A145},
          doi = {10.1051/0004-6361/202039483},
archivePrefix = {arXiv},
       eprint = {2012.07667},
 primaryClass = {astro-ph.GA},
       adsurl = {https://ui.adsabs.harvard.edu/abs/2021A&A...645A.145G}
}

@ARTICLE{Garufi2022,
       author = {{Garufi}, A. and {Podio}, L. and {Codella}, C. and {Segura-Cox}, D. and {Vander Donckt}, M. and {Mercimek}, S. and {Bacciotti}, F. and {Fedele}, D. and {Kasper}, M. and {Pineda}, J.~E. and {Humphreys}, E. and {Testi}, L.},
        title = "{ALMA chemical survey of disk-outflow sources in Taurus (ALMA-DOT). VI. Accretion shocks in the disk of DG Tau and HL Tau}",
      journal = {\aap},
         year = 2022,
        month = feb,
       volume = {658},
          eid = {A104},
        pages = {A104},
          doi = {10.1051/0004-6361/202141264},
archivePrefix = {arXiv},
       eprint = {2110.13820},
 primaryClass = {astro-ph.GA},
       adsurl = {https://ui.adsabs.harvard.edu/abs/2022A&A...658A.104G}
}

@ARTICLE{Gasman2023,
       author = {{Gasman}, Danny and {van Dishoeck}, Ewine F. and {Grant}, Sierra L. and {Temmink}, Milou and {Tabone}, Beno{\^\i}t and {Henning}, Thomas and {Kamp}, Inga and {G{\"u}del}, Manuel and {Lagage}, Pierre-Olivier and {Perotti}, Giulia and {Christiaens}, Valentin and {Samland}, Matthias and {Arabhavi}, Aditya M. and {Argyriou}, Ioannis and {Abergel}, Alain and {Absil}, Olivier and {Barrado}, David and {Boccaletti}, Anthony and {Bouwman}, Jeroen and {Caratti o Garatti}, Alessio and {Geers}, Vincent and {Glauser}, Adrian M. and {Guadarrama}, Rodrigo and {Jang}, Hyerin and {Kanwar}, Jayatee and {Lahuis}, Fred and {Morales-Calder{\'o}n}, Maria and {Mueller}, Michael and {Nehm{\'e}}, Cyrine and {Olofsson}, G{\"o}ran and {Pantin}, {\'E}ric and {Pawellek}, Nicole and {Ray}, Tom P. and {Rodgers-Lee}, Donna and {Scheithauer}, Silvia and {Schreiber}, J{\"u}rgen and {Schwarz}, Kamber and {Vandenbussche}, Bart and {Vlasblom}, Marissa and {Waters}, Rens L.~B.~F.~M. and {Wright}, Gillian and {Colina}, Luis and {Greve}, Thomas R. and {{\"O}stlin}, G{\"o}ran},
        title = "{MINDS. Abundant water and varying C/O across the disk of Sz 98 as seen by JWST/MIRI}",
      journal = {\aap},
         year = 2023,
        month = nov,
       volume = {679},
          eid = {A117},
        pages = {A117},
          doi = {10.1051/0004-6361/202347005},
archivePrefix = {arXiv},
       eprint = {2307.09301},
 primaryClass = {astro-ph.EP},
       adsurl = {https://ui.adsabs.harvard.edu/abs/2023A&A...679A.117G}
}

@ARTICLE{Goldsmith1999,
       author = {{Goldsmith}, Paul F. and {Langer}, William D.},
        title = "{Population Diagram Analysis of Molecular Line Emission}",
      journal = {\apj},
         year = 1999,
        month = may,
       volume = {517},
       number = {1},
        pages = {209-225},
          doi = {10.1086/307195},
       adsurl = {https://ui.adsabs.harvard.edu/abs/1999ApJ...517..209G}
}

@ARTICLE{Gray2016,
       author = {{Gray}, M.~D. and {Baudry}, A. and {Richards}, A.~M.~S. and {Humphreys}, E.~M.~L. and {Sobolev}, A.~M. and {Yates}, J.~A.},
        title = "{The physics of water masers observable with ALMA and SOFIA: model predictions for evolved stars}",
      journal = {\mnras},
         year = 2016,
        month = feb,
       volume = {456},
       number = {1},
        pages = {374-404},
          doi = {10.1093/mnras/stv2437},
archivePrefix = {arXiv},
       eprint = {1510.06182},
 primaryClass = {astro-ph.GA},
       adsurl = {https://ui.adsabs.harvard.edu/abs/2016MNRAS.456..374G}
}

@ARTICLE{GuerraAlvarado2024,
       author = {{Guerra-Alvarado}, Osmar M. and {Carrasco-Gonz{\'a}lez}, Carlos and {Mac{\'\i}as}, Enrique and {van der Marel}, Nienke and {Houge}, Adrien and {Maud}, Luke T. and {Pinilla}, Paola and {Villenave}, Marion and {Asaki}, Yoshiharu and {Humphreys}, Elizabeth},
        title = "{Into the thick of it: ALMA 0.45 mm observations of HL Tau at a resolution of 2 au}",
      journal = {\aap},
         year = 2024,
        month = jun,
       volume = {686},
          eid = {A298},
        pages = {A298},
          doi = {10.1051/0004-6361/202349046},
archivePrefix = {arXiv},
       eprint = {2404.04164},
 primaryClass = {astro-ph.EP},
       adsurl = {https://ui.adsabs.harvard.edu/abs/2024A&A...686A.298G}
}

@ARTICLE{Gupta2024,
       author = {{Gupta}, Aashish and {Miotello}, Anna and {Williams}, Jonathan P. and {Birnstiel}, Til and {Kuffmeier}, Michael and {Yen}, Hsi-Wei},
        title = "{TIPSY: Trajectory of Infalling Particles in Streamers around Young stars. Dynamical analysis of the streamers around S CrA and HL Tau}",
      journal = {\aap},
         year = 2024,
        month = mar,
       volume = {683},
          eid = {A133},
        pages = {A133},
          doi = {10.1051/0004-6361/202348007},
archivePrefix = {arXiv},
       eprint = {2401.10403},
 primaryClass = {astro-ph.SR},
       adsurl = {https://ui.adsabs.harvard.edu/abs/2024A&A...683A.133G}
}

@ARTICLE{Harsono2015,
       author = {{Harsono}, D. and {Bruderer}, S. and {van Dishoeck}, E.~F.},
        title = "{Volatile snowlines in embedded disks around low-mass protostars}",
      journal = {\aap},
         year = 2015,
        month = oct,
       volume = {582},
          eid = {A41},
        pages = {A41},
          doi = {10.1051/0004-6361/201525966},
archivePrefix = {arXiv},
       eprint = {1507.07480},
 primaryClass = {astro-ph.SR},
       adsurl = {https://ui.adsabs.harvard.edu/abs/2015A&A...582A..41H}
}

@ARTICLE{Hartmann1998,
       author = {{Hartmann}, Lee and {Calvet}, Nuria and {Gullbring}, Erik and {D'Alessio}, Paola},
        title = "{Accretion and the Evolution of T Tauri Disks}",
      journal = {\apj},
         year = 1998,
        month = mar,
       volume = {495},
       number = {1},
        pages = {385-400},
          doi = {10.1086/305277},
       adsurl = {https://ui.adsabs.harvard.edu/abs/1998ApJ...495..385H}
}

@ARTICLE{Houge2025,
       author = {{Houge}, Adrien and {Krijt}, Sebastiaan and {Banzatti}, Andrea and {Blake}, Geoffrey A and {Pinilla}, Paola and {Pontoppidan}, Klaus M and {Trapman}, Leon and {Williams}, Joe and {Zhang}, Ke},
        title = "{Smuggling unnoticed: towards a 2D view of water and dust delivery to the inner regions of protoplanetary discs}",
      journal = {\mnras},
         year = 2025,
        month = feb,
       volume = {537},
       number = {2},
        pages = {691-704},
          doi = {10.1093/mnras/staf057},
archivePrefix = {arXiv},
       eprint = {2501.05881},
 primaryClass = {astro-ph.EP},
       adsurl = {https://ui.adsabs.harvard.edu/abs/2025MNRAS.537..691H}
}

@ARTICLE{Isella2016,
       author = {{Isella}, Andrea and {Guidi}, Greta and {Testi}, Leonardo and {Liu}, Shangfei and {Li}, Hui and {Li}, Shengtai and {Weaver}, Erik and {Boehler}, Yann and {Carperter}, John M. and {De Gregorio-Monsalvo}, Itziar and {Manara}, Carlo F. and {Natta}, Antonella and {P{\'e}rez}, Laura M. and {Ricci}, Luca and {Sargent}, Anneila and {Tazzari}, Marco and {Turner}, Neal},
        title = "{Ringed Structures of the HD 163296 Protoplanetary Disk Revealed by ALMA}",
      journal = {\prl},
         year = 2016,
        month = dec,
       volume = {117},
       number = {25},
          eid = {251101},
        pages = {251101},
          doi = {10.1103/PhysRevLett.117.251101},
       adsurl = {https://ui.adsabs.harvard.edu/abs/2016PhRvL.117y1101I}
}

@ARTICLE{JvM1995,
       author = {{Jorsater}, Steven and {van Moorsel}, Gustaaf A.},
        title = "{High Resolution Neutral Hydrogen Observations of the Barred Spiral Galaxy NGC 1365}",
      journal = {\aj},
         year = 1995,
        month = nov,
       volume = {110},
        pages = {2037},
          doi = {10.1086/117668},
       adsurl = {https://ui.adsabs.harvard.edu/abs/1995AJ....110.2037J}
}

@ARTICLE{Kaeufer2024,
       author = {{Kaeufer}, T. and {Min}, M. and {Woitke}, P. and {Kamp}, I. and {Arabhavi}, A.~M.},
        title = "{Bayesian analysis of the molecular emission and dust continuum of protoplanetary disks}",
      journal = {\aap},
         year = 2024,
        month = jul,
       volume = {687},
          eid = {A209},
        pages = {A209},
          doi = {10.1051/0004-6361/202449936},
archivePrefix = {arXiv},
       eprint = {2405.06486},
 primaryClass = {astro-ph.EP},
       adsurl = {https://ui.adsabs.harvard.edu/abs/2024A&A...687A.209K}
}

@ARTICLE{Kristensen2016,
       author = {{Kristensen}, Lars E. and {Brown}, Joanna M. and {Wilner}, David and {Salyk}, Colette},
        title = "{Velocity-resolved Hot Water Emission Detected toward HL Tau with the Submillimeter Array}",
      journal = {\apjl},
         year = 2016,
        month = may,
       volume = {822},
       number = {1},
          eid = {L20},
        pages = {L20},
          doi = {10.3847/2041-8205/822/1/L20},
archivePrefix = {arXiv},
       eprint = {1603.09594},
 primaryClass = {astro-ph.SR},
       adsurl = {https://ui.adsabs.harvard.edu/abs/2016ApJ...822L..20K}
}

@ARTICLE{Leemker2021,
       author = {{Leemker}, M. and {van't Hoff}, M.~L.~R. and {Trapman}, L. and {van Gelder}, M.~L. and {Hogerheijde}, M.~R. and {Ru{\'\i}z-Rodr{\'\i}guez}, D. and {van Dishoeck}, E.~F.},
        title = "{Chemically tracing the water snowline in protoplanetary disks with HCO$^{+}$}",
      journal = {\aap},
         year = 2021,
        month = feb,
       volume = {646},
          eid = {A3},
        pages = {A3},
          doi = {10.1051/0004-6361/202039387},
archivePrefix = {arXiv},
       eprint = {2011.12319},
 primaryClass = {astro-ph.EP},
       adsurl = {https://ui.adsabs.harvard.edu/abs/2021A&A...646A...3L}
}

@ARTICLE{Leemker2022,
       author = {{Leemker}, M. and {Booth}, A.~S. and {van Dishoeck}, E.~F. and {P{\'e}rez-S{\'a}nchez}, A.~F. and {Szul{\'a}gyi}, J. and {Bosman}, A.~D. and {Bruderer}, S. and {Facchini}, S. and {Hogerheijde}, M.~R. and {Paneque-Carre{\~n}o}, T. and {Sturm}, J.~A.},
        title = "{Gas temperature structure across transition disk cavities}",
      journal = {\aap},
         year = 2022,
        month = jul,
       volume = {663},
          eid = {A23},
        pages = {A23},
          doi = {10.1051/0004-6361/202243229},
archivePrefix = {arXiv},
       eprint = {2204.03666},
 primaryClass = {astro-ph.EP},
       adsurl = {https://ui.adsabs.harvard.edu/abs/2022A&A...663A..23L}
}

@ARTICLE{Leemker2023,
       author = {{Leemker}, M. and {Booth}, A.~S. and {van Dishoeck}, E.~F. and {van der Marel}, N. and {Tabone}, B. and {Ligterink}, N.~F.~W. and {Brunken}, N.~G.~C. and {Hogerheijde}, M.~R.},
        title = "{A major asymmetric ice trap in a planet-forming disk. IV. Nitric oxide gas and a lack of CN tracing sublimating ices and a C/O ratio <1}",
      journal = {\aap},
         year = 2023,
        month = may,
       volume = {673},
          eid = {A7},
        pages = {A7},
          doi = {10.1051/0004-6361/202245662},
archivePrefix = {arXiv},
       eprint = {2303.00768},
 primaryClass = {astro-ph.EP},
       adsurl = {https://ui.adsabs.harvard.edu/abs/2023A&A...673A...7L}
}

@ARTICLE{Liu2017,
       author = {{Liu}, Yao and {Henning}, Thomas and {Carrasco-Gonz{\'a}lez}, Carlos and {Chandler}, Claire J. and {Linz}, Hendrik and {Birnstiel}, Til and {van Boekel}, Roy and {P{\'e}rez}, Laura M. and {Flock}, Mario and {Testi}, Leonardo and {Rodr{\'\i}guez}, Luis F. and {Galv{\'a}n-Madrid}, Roberto},
        title = "{The properties of the inner disk around HL Tau: Multi-wavelength modeling of the dust emission}",
      journal = {\aap},
         year = 2017,
        month = nov,
       volume = {607},
          eid = {A74},
        pages = {A74},
          doi = {10.1051/0004-6361/201629786},
archivePrefix = {arXiv},
       eprint = {1708.03238},
 primaryClass = {astro-ph.EP},
       adsurl = {https://ui.adsabs.harvard.edu/abs/2017A&A...607A..74L}
}

@ARTICLE{Loomis2018,
       author = {{Loomis}, Ryan A. and {Cleeves}, L. Ilsedore and {{\"O}berg}, Karin I. and {Aikawa}, Yuri and {Bergner}, Jennifer and {Furuya}, Kenji and {Guzman}, V.~V. and {Walsh}, Catherine},
        title = "{The Distribution and Excitation of CH$_{3}$CN in a Solar Nebula Analog}",
      journal = {\apj},
         year = 2018,
        month = jun,
       volume = {859},
       number = {2},
          eid = {131},
        pages = {131},
          doi = {10.3847/1538-4357/aac169},
archivePrefix = {arXiv},
       eprint = {1805.01458},
 primaryClass = {astro-ph.EP},
       adsurl = {https://ui.adsabs.harvard.edu/abs/2018ApJ...859..131L}
}

@ARTICLE{Loomis2025,
       author = {{Loomis}, Ryan A. and {Facchini}, Stefano and {Benisty}, Myriam and {Curone}, Pietro and {Ilee}, John D. and {Cataldi}, Gianni and {Yen}, Hsi-Wei and {Teague}, Richard and {Pinte}, Christophe and {Huang}, Jane and {Garg}, Himanshi and {Orihara}, Ryuta and {Czekala}, Ian and {Zawadzki}, Brianna and {Andrews}, Sean M. and {Wilner}, David J. and {Bae}, Jaehan and {Barraza-Alfaro}, Marcelo and {Fasano}, Daniele and {Flock}, Mario and {Fukagawa}, Misato and {Galloway-Sprietsma}, Maria and {Izquierdo}, Andr{\'e}s F. and {Kanagawa}, Kazuhiro and {Lesur}, Geoffroy and {Longarini}, Cristiano and {Menard}, Francois and {Price}, Daniel J. and {Rosotti}, Giovanni and {Stadler}, Jochen and {Wafflard-Fernandez}, Gaylor and {W{\"o}lfer}, Lisa and {Yoshida}, Tomohiro C.},
        title = "{exoALMA. II. Data Calibration and Imaging Pipeline}",
      journal = {\apjl},
         year = 2025,
        month = may,
       volume = {984},
       number = {1},
          eid = {L7},
        pages = {L7},
          doi = {10.3847/2041-8213/adc43a},
archivePrefix = {arXiv},
       eprint = {2504.19870},
 primaryClass = {astro-ph.EP},
       adsurl = {https://ui.adsabs.harvard.edu/abs/2025ApJ...984L...7L}
}

@ARTICLE{Lumbreras2014,
       author = {{Lumbreras}, Alba M. and {Zapata}, Luis A.},
        title = "{SMA Submillimeter Observations of HL Tau: Revealing a Compact Molecular Outflow}",
      journal = {\aj},
         year = 2014,
        month = apr,
       volume = {147},
       number = {4},
          eid = {72},
        pages = {72},
          doi = {10.1088/0004-6256/147/4/72},
archivePrefix = {arXiv},
       eprint = {1401.0455},
 primaryClass = {astro-ph.GA},
       adsurl = {https://ui.adsabs.harvard.edu/abs/2014AJ....147...72L}
}

@ARTICLE{Lynden-Bell1974,
       author = {{Lynden-Bell}, D. and {Pringle}, J.~E.},
        title = "{The evolution of viscous discs and the origin of the nebular variables.}",
      journal = {\mnras},
         year = 1974,
        month = sep,
       volume = {168},
        pages = {603-637},
          doi = {10.1093/mnras/168.3.603},
       adsurl = {https://ui.adsabs.harvard.edu/abs/1974MNRAS.168..603L}
}

@INPROCEEDINGS{McMullin2007,
       author = {{McMullin}, J.~P. and {Waters}, B. and {Schiebel}, D. and {Young}, W. and {Golap}, K.},
        title = "{CASA Architecture and Applications}",
    booktitle = {Astronomical Data Analysis Software and Systems XVI},
         year = 2007,
       editor = {{Shaw}, R.~A. and {Hill}, F. and {Bell}, D.~J.},
       series = {Astronomical Society of the Pacific Conference Series},
       volume = {376},
        month = oct,
        pages = {127},
       adsurl = {https://ui.adsabs.harvard.edu/abs/2007ASPC..376..127M}
}

@ARTICLE{Miotello2016,
       author = {{Miotello}, A. and {van Dishoeck}, E.~F. and {Kama}, M. and {Bruderer}, S.},
        title = "{Determining protoplanetary disk gas masses from CO isotopologues line observations}",
      journal = {\aap},
         year = 2016,
        month = oct,
       volume = {594},
          eid = {A85},
        pages = {A85},
          doi = {10.1051/0004-6361/201628159},
archivePrefix = {arXiv},
       eprint = {1605.07780},
 primaryClass = {astro-ph.SR},
       adsurl = {https://ui.adsabs.harvard.edu/abs/2016A&A...594A..85M}
}

@ARTICLE{Milam2005,
       author = {{Milam}, S.~N. and {Savage}, C. and {Brewster}, M.~A. and {Ziurys}, L.~M. and {Wyckoff}, S.},
        title = "{The $^{12}$C/$^{13}$C Isotope Gradient Derived from Millimeter Transitions of CN: The Case for Galactic Chemical Evolution}",
      journal = {\apj},
         year = 2005,
        month = dec,
       volume = {634},
       number = {2},
        pages = {1126-1132},
          doi = {10.1086/497123},
       adsurl = {https://ui.adsabs.harvard.edu/abs/2005ApJ...634.1126M}
}

@ARTICLE{Oberg2011,
       author = {{{\"O}berg}, Karin I. and {Murray-Clay}, Ruth and {Bergin}, Edwin A.},
        title = "{The Effects of Snowlines on C/O in Planetary Atmospheres}",
      journal = {\apjl},
         year = 2011,
        month = dec,
       volume = {743},
       number = {1},
          eid = {L16},
        pages = {L16},
          doi = {10.1088/2041-8205/743/1/L16},
archivePrefix = {arXiv},
       eprint = {1110.5567},
 primaryClass = {astro-ph.GA},
       adsurl = {https://ui.adsabs.harvard.edu/abs/2011ApJ...743L..16O}
}

@ARTICLE{Oberg2021,
       author = {{{\"O}berg}, Karin I. and {Bergin}, Edwin A.},
        title = "{Astrochemistry and compositions of planetary systems}",
      journal = {\physrep},
         year = 2021,
        month = jan,
       volume = {893},
        pages = {1-48},
          doi = {10.1016/j.physrep.2020.09.004},
archivePrefix = {arXiv},
       eprint = {2010.03529},
 primaryClass = {astro-ph.EP},
       adsurl = {https://ui.adsabs.harvard.edu/abs/2021PhR...893....1O}
}

@ARTICLE{Phillips1992,
       author = {{Phillips}, T.~G. and {van Dishoeck}, Ewine F. and {Keene}, Jocelyn},
        title = "{Interstellar H 3O + and Its Relation to the O 2 and H 2O Abundances}",
      journal = {\apj},
         year = 1992,
        month = nov,
       volume = {399},
        pages = {533},
          doi = {10.1086/171945},
       adsurl = {https://ui.adsabs.harvard.edu/abs/1992ApJ...399..533P}
}

@ARTICLE{Pickett1998,
       author = {{Pickett}, H.~M. and {Poynter}, R.~L. and {Cohen}, E.~A. and {Delitsky}, M.~L. and {Pearson}, J.~C. and {M{\"u}ller}, H.~S.~P.},
        title = "{Submillimeter, millimeter and microwave spectral line catalog.}",
      journal = {\jqsrt},
         year = 1998,
        month = nov,
       volume = {60},
       number = {5},
        pages = {883-890},
          doi = {10.1016/S0022-4073(98)00091-0},
       adsurl = {https://ui.adsabs.harvard.edu/abs/1998JQSRT..60..883P}
}

@ARTICLE{Pinte2016,
       author = {{Pinte}, C. and {Dent}, W.~R.~F. and {M{\'e}nard}, F. and {Hales}, A. and {Hill}, T. and {Cortes}, P. and {de Gregorio-Monsalvo}, I.},
        title = "{Dust and Gas in the Disk of HL Tauri: Surface Density, Dust Settling, and Dust-to-gas Ratio}",
      journal = {\apj},
         year = 2016,
        month = jan,
       volume = {816},
       number = {1},
          eid = {25},
        pages = {25},
          doi = {10.3847/0004-637X/816/1/25},
archivePrefix = {arXiv},
       eprint = {1508.00584},
 primaryClass = {astro-ph.SR},
       adsurl = {https://ui.adsabs.harvard.edu/abs/2016ApJ...816...25P}
}

@ARTICLE{Pirovano2022,
       author = {{Pirovano}, L.~M. and {Fedele}, D. and {van Dishoeck}, E.~F. and {Hogerheijde}, M.~R. and {Lodato}, G. and {Bruderer}, S.},
        title = "{H$_{2}$O distribution in the disc of HD 100546 and HD 163296: the role of dust dynamics and planet-disc interaction}",
      journal = {\aap},
         year = 2022,
        month = sep,
       volume = {665},
          eid = {A45},
        pages = {A45},
          doi = {10.1051/0004-6361/202244104},
archivePrefix = {arXiv},
       eprint = {2207.10744},
 primaryClass = {astro-ph.EP},
       adsurl = {https://ui.adsabs.harvard.edu/abs/2022A&A...665A..45P}
}

@ARTICLE{Pontoppidan2010,
       author = {{Pontoppidan}, Klaus M. and {Salyk}, Colette and {Blake}, Geoffrey A. and {K{\"a}ufl}, Hans Ulrich},
        title = "{Spectrally Resolved Pure Rotational Lines of Water in Protoplanetary Disks}",
      journal = {\apjl},
         year = 2010,
        month = oct,
       volume = {722},
       number = {2},
        pages = {L173-L177},
          doi = {10.1088/2041-8205/722/2/L173},
archivePrefix = {arXiv},
       eprint = {1009.3259},
 primaryClass = {astro-ph.EP},
       adsurl = {https://ui.adsabs.harvard.edu/abs/2010ApJ...722L.173P}
}

@ARTICLE{Pontoppidan2024,
       author = {{Pontoppidan}, Klaus M. and {Salyk}, Colette and {Banzatti}, Andrea and {Zhang}, Ke and {Pascucci}, Ilaria and {{\"O}berg}, Karin I. and {Long}, Feng and {Romero-Mirza}, Carlos E. and {Carr}, John and {Najita}, Joan and {Blake}, Geoffrey A. and {Arulanantham}, Nicole and {Andrews}, Sean and {Ballering}, Nicholas P. and {Bergin}, Edwin and {Calahan}, Jenny and {Cobb}, Douglas and {Colmenares}, Maria Jose and {Dickson-Vandervelde}, Annie and {Dignan}, Anna and {Green}, Joel and {Heretz}, Phoebe and {Herczeg}, Gregory and {Kalyaan}, Anusha and {Krijt}, Sebastiaan and {Pauly}, Tyler and {Pinilla}, Paola and {Trapman}, Leon and {Xie}, Chengyan},
        title = "{High-contrast JWST-MIRI Spectroscopy of Planet-forming Disks for the JDISC Survey}",
      journal = {\apj},
         year = 2024,
        month = mar,
       volume = {963},
       number = {2},
          eid = {158},
        pages = {158},
          doi = {10.3847/1538-4357/ad20f0},
archivePrefix = {arXiv},
       eprint = {2311.17020},
 primaryClass = {astro-ph.EP},
       adsurl = {https://ui.adsabs.harvard.edu/abs/2024ApJ...963..158P}
}

@ARTICLE{Rampinelliinprep,
       author = {{Rampinelli}, Luna and {Facchini}, Stefano and {Leemker}, Margot and {Isella}, Andrea and {Curone}, Pietro and {Benisty}, Myriam and {Humphreys}, Elizabeth and {Testi}, Leonardo},
        title = "{Water Vapor Emission at the Warm Cavity Wall of the HD 100546 Disk as Revealed by ALMA}",
      journal = {\apjl},
         year = 2026,
        month = jan,
       volume = {996},
       number = {1},
          eid = {L17},
        pages = {L17},
          doi = {10.3847/2041-8213/ae2868},
archivePrefix = {arXiv},
       eprint = {2512.06439},
 primaryClass = {astro-ph.EP},
       adsurl = {https://ui.adsabs.harvard.edu/abs/2026ApJ...996L..17R}
}

@ARTICLE{Rebull2004,
       author = {{Rebull}, L.~M. and {Wolff}, S.~C. and {Strom}, S.~E.},
        title = "{Stellar Rotation in Young Clusters: The First 4 Million Years}",
      journal = {\aj},
         year = 2004,
        month = feb,
       volume = {127},
       number = {2},
        pages = {1029-1051},
          doi = {10.1086/380931},
       adsurl = {https://ui.adsabs.harvard.edu/abs/2004AJ....127.1029R}
}

@ARTICLE{RiviereMarichalar2012,
       author = {{Riviere-Marichalar}, P. and {M{\'e}nard}, F. and {Thi}, W.~F. and {Kamp}, I. and {Montesinos}, B. and {Meeus}, G. and {Woitke}, P. and {Howard}, C. and {Sandell}, G. and {Podio}, L. and {Dent}, W.~R.~F. and {Mendigut{\'\i}a}, I. and {Pinte}, C. and {White}, G.~J. and {Barrado}, D.},
        title = "{Detection of warm water vapour in Taurus protoplanetary discs by Herschel}",
      journal = {\aap},
         year = 2012,
        month = feb,
       volume = {538},
          eid = {L3},
        pages = {L3},
          doi = {10.1051/0004-6361/201118448},
archivePrefix = {arXiv},
       eprint = {1112.4850},
 primaryClass = {astro-ph.SR},
       adsurl = {https://ui.adsabs.harvard.edu/abs/2012A&A...538L...3R}
}

@ARTICLE{Robitaille2007,
       author = {{Robitaille}, Thomas P. and {Whitney}, Barbara A. and {Indebetouw}, Remy and {Wood}, Kenneth},
        title = "{Interpreting Spectral Energy Distributions from Young Stellar Objects. II. Fitting Observed SEDs Using a Large Grid of Precomputed Models}",
      journal = {\apjs},
         year = 2007,
        month = apr,
       volume = {169},
       number = {2},
        pages = {328-352},
          doi = {10.1086/512039},
archivePrefix = {arXiv},
       eprint = {astro-ph/0612690},
 primaryClass = {astro-ph},
       adsurl = {https://ui.adsabs.harvard.edu/abs/2007ApJS..169..328R}
}

@ARTICLE{Salyk2019,
       author = {{Salyk}, Colette and {Lacy}, John and {Richter}, Matt and {Zhang}, Ke and {Pontoppidan}, Klaus and {Carr}, John S. and {Najita}, Joan R. and {Blake}, Geoffrey A.},
        title = "{A High-resolution Mid-infrared Survey of Water Emission from Protoplanetary Disks}",
      journal = {\apj},
         year = 2019,
        month = mar,
       volume = {874},
       number = {1},
          eid = {24},
        pages = {24},
          doi = {10.3847/1538-4357/ab05c3},
archivePrefix = {arXiv},
       eprint = {1902.02708},
 primaryClass = {astro-ph.SR},
       adsurl = {https://ui.adsabs.harvard.edu/abs/2019ApJ...874...24S}
}

@ARTICLE{Schoier2005,
       author = {{Sch{\"o}ier}, F.~L. and {van der Tak}, F.~F.~S. and {van Dishoeck}, E.~F. and {Black}, J.~H.},
        title = "{An atomic and molecular database for analysis of submillimetre line observations}",
      journal = {\aap},
         year = 2005,
        month = mar,
       volume = {432},
       number = {1},
        pages = {369-379},
          doi = {10.1051/0004-6361:20041729},
archivePrefix = {arXiv},
       eprint = {astro-ph/0411110},
 primaryClass = {astro-ph},
       adsurl = {https://ui.adsabs.harvard.edu/abs/2005A&A...432..369S}
}

@ARTICLE{Schoonenberg2017,
       author = {{Schoonenberg}, Djoeke and {Ormel}, Chris W.},
        title = "{Planetesimal formation near the snowline: in or out?}",
      journal = {\aap},
         year = 2017,
        month = jun,
       volume = {602},
          eid = {A21},
        pages = {A21},
          doi = {10.1051/0004-6361/201630013},
archivePrefix = {arXiv},
       eprint = {1702.02151},
 primaryClass = {astro-ph.EP},
       adsurl = {https://ui.adsabs.harvard.edu/abs/2017A&A...602A..21S}
}

@ARTICLE{Skinner2020,
       author = {{Skinner}, Stephen L. and {G{\"u}del}, Manuel},
        title = "{X-Ray Emission and Disk Irradiation of HL Tau and HD 100546}",
      journal = {\apj},
         year = 2020,
        month = jan,
       volume = {888},
       number = {1},
          eid = {15},
        pages = {15},
          doi = {10.3847/1538-4357/ab585c},
       adsurl = {https://ui.adsabs.harvard.edu/abs/2020ApJ...888...15S}
}

@ARTICLE{Soaveinprep,
       author = {{Soave}, Alessandro and {Leemker}, M. and {Facchini}, S. and {Testi}, L. },
        title = "",
      journal = {\aap},
         year = 2025,
        month = dec,
       volume = {in prep.},
       number = {},
          eid = {},
        pages = {},
          doi = {},
       adsurl = {}
}

@ARTICLE{Teague2019,
       author = {{Teague}, Richard},
        title = "{GoFish: Fishing for Line Observations in Protoplanetary Disks}",
      journal = {The Journal of Open Source Software},
         year = 2019,
        month = sep,
       volume = {4},
       number = {41},
          eid = {1632},
        pages = {1632},
          doi = {10.21105/joss.01632},
       adsurl = {https://ui.adsabs.harvard.edu/abs/2019JOSS....4.1632T}
}

@ARTICLE{Temmink2024,
       author = {{Temmink}, Milou and {van Dishoeck}, Ewine F. and {Gasman}, Danny and {Grant}, Sierra L. and {Tabone}, Beno{\^\i}t and {G{\"u}del}, Manuel and {Henning}, Thomas and {Barrado}, David and {Caratti o Garatti}, Alessio and {Glauser}, Adrian M. and {Kamp}, Inga and {Arabhavi}, Aditya M. and {Jang}, Hyerin and {Kurtovic}, Nicolas and {Perotti}, Giulia and {Schwarz}, Kamber and {Vlasblom}, Marissa},
        title = "{MINDS: The DR Tau disk: II. Probing the hot and cold H$_{2}$O reservoirs in the JWST-MIRI spectrum}",
      journal = {\aap},
         year = 2024,
        month = sep,
       volume = {689},
          eid = {A330},
        pages = {A330},
          doi = {10.1051/0004-6361/202450355},
archivePrefix = {arXiv},
       eprint = {2407.05070},
 primaryClass = {astro-ph.EP},
       adsurl = {https://ui.adsabs.harvard.edu/abs/2024A&A...689A.330T}
}

@ARTICLE{Tennyson2001,
       author = {{Tennyson}, Jonathan and {Zobov}, Nikolai F. and {Williamson}, Ross and {Polyansky}, Oleg L. and {Bernath}, Peter F.},
        title = "{Experimental Energy Levels of the Water Molecule}",
      journal = {Journal of Physical and Chemical Reference Data},
         year = 2001,
        month = may,
       volume = {30},
       number = {3},
        pages = {735-831},
          doi = {10.1063/1.1364517},
       adsurl = {https://ui.adsabs.harvard.edu/abs/2001JPCRD..30..735T}
}

@ARTICLE{Tobin2023,
       author = {{Tobin}, John J. and {van't Hoff}, Merel L.~R. and {Leemker}, Margot and {van Dishoeck}, Ewine F. and {Paneque-Carre{\~n}o}, Teresa and {Furuya}, Kenji and {Harsono}, Daniel and {Persson}, Magnus V. and {Cleeves}, L. Ilsedore and {Sheehan}, Patrick D. and {Cieza}, Lucas},
        title = "{Deuterium-enriched water ties planet-forming disks to comets and protostars}",
      journal = {\nat},
         year = 2023,
        month = mar,
       volume = {615},
       number = {7951},
        pages = {227-230},
          doi = {10.1038/s41586-022-05676-z},
       adsurl = {https://ui.adsabs.harvard.edu/abs/2023Natur.615..227T}
}

@ARTICLE{TorresVillanuevainprep,
       author = {{TorresVillanueva}, E. and {et al.}},
        title = "{}",
      journal = {},
         year = 2026,
        month = dec,
       volume = {in prep.},
       number = {},
        pages = {},
          doi = {},
       adsurl = {}
}

@ARTICLE{Ueda2025,
       author = {{Ueda}, Takahiro and {Andrews}, Sean M. and {Carrasco-Gonz{\'a}lez}, Carlos and {Guerra-Alvarado}, Osmar M. and {Okuzumi}, Satoshi and {Tazaki}, Ryo and {Kataoka}, Akimasa},
        title = "{Multi-Wavelength Dust Characterization of the HL Tau Disk and Implications for Planet Formation}",
      journal = {arXiv e-prints},
         year = 2025,
        month = jul,
          eid = {arXiv:2507.14443},
        pages = {arXiv:2507.14443},
          doi = {10.48550/arXiv.2507.14443},
archivePrefix = {arXiv},
       eprint = {2507.14443},
 primaryClass = {astro-ph.EP},
       adsurl = {https://ui.adsabs.harvard.edu/abs/2025arXiv250714443U}
}

@ARTICLE{vanDishoeck2021,
       author = {{van Dishoeck}, E.~F. and {Kristensen}, L.~E. and {Mottram}, J.~C. and {Benz}, A.~O. and {Bergin}, E.~A. and {Caselli}, P. and {Herpin}, F. and {Hogerheijde}, M.~R. and {Johnstone}, D. and {Liseau}, R. and {Nisini}, B. and {Tafalla}, M. and {van der Tak}, F.~F.~S. and {Wyrowski}, F. and {Baudry}, A. and {Benedettini}, M. and {Bjerkeli}, P. and {Blake}, G.~A. and {Braine}, J. and {Bruderer}, S. and {Cabrit}, S. and {Cernicharo}, J. and {Choi}, Y. and {Coutens}, A. and {de Graauw}, Th. and {Dominik}, C. and {Fedele}, D. and {Fich}, M. and {Fuente}, A. and {Furuya}, K. and {Goicoechea}, J.~R. and {Harsono}, D. and {Helmich}, F.~P. and {Herczeg}, G.~J. and {Jacq}, T. and {Karska}, A. and {Kaufman}, M. and {Keto}, E. and {Lamberts}, T. and {Larsson}, B. and {Leurini}, S. and {Lis}, D.~C. and {Melnick}, G. and {Neufeld}, D. and {Pagani}, L. and {Persson}, M. and {Shipman}, R. and {Taquet}, V. and {van Kempen}, T.~A. and {Walsh}, C. and {Wampfler}, S.~F. and {Y{\i}ld{\i}z}, U. and {WISH Team}},
        title = "{Water in star-forming regions: physics and chemistry from clouds to disks as probed by Herschel spectroscopy}",
      journal = {\aap},
         year = 2021,
        month = apr,
       volume = {648},
          eid = {A24},
        pages = {A24},
          doi = {10.1051/0004-6361/202039084},
archivePrefix = {arXiv},
       eprint = {2102.02225},
 primaryClass = {astro-ph.GA},
       adsurl = {https://ui.adsabs.harvard.edu/abs/2021A&A...648A..24V}
}

@ARTICLE{vantHoff2018,
       author = {{van 't Hoff}, Merel L.~R. and {Tobin}, John J. and {Trapman}, Leon and {Harsono}, Daniel and {Sheehan}, Patrick D. and {Fischer}, William J. and {Megeath}, S. Thomas and {van Dishoeck}, Ewine F.},
        title = "{Methanol and its Relation to the Water Snowline in the Disk around the Young Outbursting Star V883 Ori}",
      journal = {\apjl},
         year = 2018,
        month = sep,
       volume = {864},
       number = {1},
          eid = {L23},
        pages = {L23},
          doi = {10.3847/2041-8213/aadb8a},
archivePrefix = {arXiv},
       eprint = {1808.08258},
 primaryClass = {astro-ph.SR},
       adsurl = {https://ui.adsabs.harvard.edu/abs/2018ApJ...864L..23V}
}

@ARTICLE{Vlasblom2025,
       author = {{Vlasblom}, Marissa and {Temmink}, Milou and {Sellek}, Andrew D. and {van Dishoeck}, Ewine F.},
        title = "{Understanding JWST water spectra: What can thermochemical models tell us about the (cold) water in protoplanetary disks?}",
      journal = {\aap},
         year = 2025,
        month = nov,
       volume = {703},
          eid = {A52},
        pages = {A52},
          doi = {10.1051/0004-6361/202555809},
archivePrefix = {arXiv},
       eprint = {2509.06494},
 primaryClass = {astro-ph.EP},
       adsurl = {https://ui.adsabs.harvard.edu/abs/2025A&A...703A..52V}
}

@ARTICLE{Wang2025,
       author = {{Wang}, Yu and {Ormel}, Chris W. and {Mori}, Shoji and {Bai}, Xue-Ning},
        title = "{Solving for the 2D water snowline with hydrodynamic simulations: Emergence of the gas outflow, water cycle, and temperature plateau}",
      journal = {\aap},
         year = 2025,
        month = apr,
       volume = {696},
          eid = {A38},
        pages = {A38},
          doi = {10.1051/0004-6361/202453036},
archivePrefix = {arXiv},
       eprint = {2502.08936},
 primaryClass = {astro-ph.EP},
       adsurl = {https://ui.adsabs.harvard.edu/abs/2025A&A...696A..38W}
}

@ARTICLE{Weaver2018,
       author = {{Weaver}, Erik and {Isella}, Andrea and {Boehler}, Yann},
        title = "{Empirical Temperature Measurement in Protoplanetary Disks}",
      journal = {\apj},
         year = 2018,
        month = feb,
       volume = {853},
       number = {2},
          eid = {113},
        pages = {113},
          doi = {10.3847/1538-4357/aaa481},
archivePrefix = {arXiv},
       eprint = {1801.03478},
 primaryClass = {astro-ph.EP},
       adsurl = {https://ui.adsabs.harvard.edu/abs/2018ApJ...853..113W}
}

@ARTICLE{Wilson1999,
       author = {{Wilson}, T.~L.},
        title = "{Isotopes in the interstellar medium and circumstellar envelopes}",
      journal = {Reports on Progress in Physics},
         year = 1999,
        month = feb,
       volume = {62},
       number = {2},
        pages = {143-185},
          doi = {10.1088/0034-4885/62/2/002},
       adsurl = {https://ui.adsabs.harvard.edu/abs/1999RPPh...62..143W}
}

@ARTICLE{Yang2025,
       author = {{Yang}, Haifeng and {Stephens}, Ian W. and {Lin}, Zhe-Yu Daniel and {Fern{\'a}ndez-L{\'o}pez}, Manuel and {Li}, Zhi-Yun and {Looney}, Leslie W. and {Harrison}, Rachel},
        title = "{Detailed radial scale height profile of dust grains as probed by dust self-scattering in HL Tau}",
      journal = {arXiv e-prints},
         year = 2025,
        month = aug,
          eid = {arXiv:2508.01233},
        pages = {arXiv:2508.01233},
          doi = {10.48550/arXiv.2508.01233},
archivePrefix = {arXiv},
       eprint = {2508.01233},
 primaryClass = {astro-ph.EP},
       adsurl = {https://ui.adsabs.harvard.edu/abs/2025arXiv250801233Y}
}

@ARTICLE{Yen2019,
       author = {{Yen}, Hsi-Wei and {Gu}, Pin-Gao and {Hirano}, Naomi and {Koch}, Patrick M. and {Lee}, Chin-Fei and {Liu}, Hauyu Baobab and {Takakuwa}, Shigehisa},
        title = "{HL Tau Disk in HCO$^{+}$ (3-2) and (1-0) with ALMA: Gas Density, Temperature, Gap, and One-arm Spiral}",
      journal = {\apj},
         year = 2019,
        month = aug,
       volume = {880},
       number = {2},
          eid = {69},
        pages = {69},
          doi = {10.3847/1538-4357/ab29f8},
archivePrefix = {arXiv},
       eprint = {1906.05535},
 primaryClass = {astro-ph.SR},
       adsurl = {https://ui.adsabs.harvard.edu/abs/2019ApJ...880...69Y}
}

@ARTICLE{Yu2012,
       author = {{Yu}, Shanshan and {Pearson}, John C. and {Drouin}, Brian J. and {Martin-Drumel}, Marie-Aline and {Pirali}, Olivier and {Vervloet}, Michel and {Coudert}, Laurent H. and {M{\"u}ller}, Holger S.~P. and {Br{\"u}nken}, Sandra},
        title = "{Measurement and analysis of new terahertz and far-infrared spectra of high temperature water}",
      journal = {Journal of Molecular Spectroscopy},
         year = 2012,
        month = sep,
       volume = {279},
        pages = {16-25},
          doi = {10.1016/j.jms.2012.07.011},
       adsurl = {https://ui.adsabs.harvard.edu/abs/2012JMoSp.279...16Y}
}

\appendix

\section{Observations}

\subsection{Frequency dependent noise in the \ce{H2O} 183~GHz spectra} \label{app:telluric}

\begin{figure*}
\centering
\includegraphics[width=\hsize]{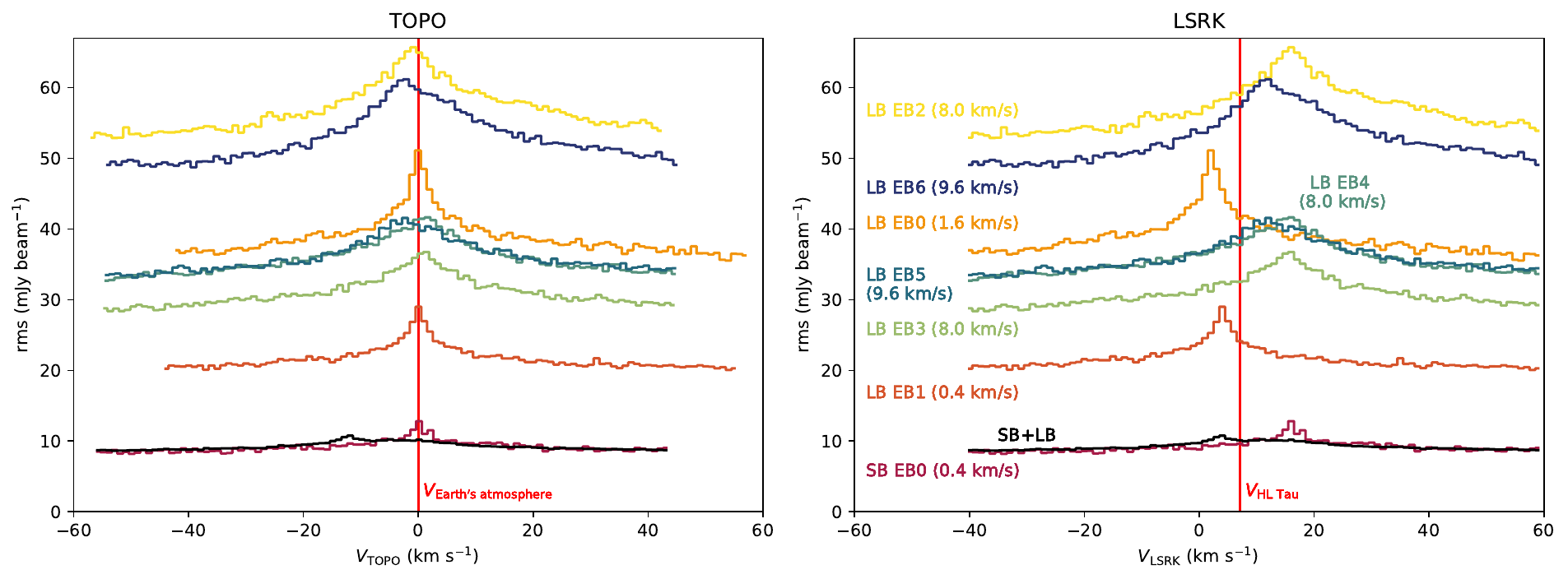}
  \caption{RMS as function of frequency in the spws covering the \ce{H2O} 183~GHz line imaged with a \texttt{robust} parameter of 0.0. The left panel shows the spectra in topocentric coordinates where the telluric due to the water in the Earth's atmosphere increases the noise at a velocity of 0~km~s$^{-1}$. The right panel shows the same spectra but then in the local standard of rest frame. The number within brackets in the labels indicates the velocity width used for the bandpass calibration. }
     \label{fig:rms_spectra}
\end{figure*}

The \ce{H2O} 183~GHz line has an upper energy level of 205~K, comparable to the temperature of the water in some layers of the Earth's atmosphere. Therefore, this line is by definition located in a strong telluric. The effect of this telluric on the noise in the spw covering the 183~GHz \ce{H2O} line is demonstrated in Fig.~\ref{fig:rms_spectra}, where the left panel presents the spectra as function of the topocentric velocity where the telluric is at 0~km~s$^{-1}$ and the right panel presents those in the local standard of rest frame. 
Each spectrum corresponds to one EB of data taken during the short or long baseline EBs in Table~\ref{tab:obs_details}. The spectra are extracted from the cubes imaged with a \texttt{robust} parameter of 0.0 from a 7" square region excluding the inner 1"$\times 0\farcs7$ elliptical region where the continuum is seen using \texttt{CARTA} \citep{carta}.
The noise in the short baseline data (purple) is extremely low due to the exceptional weather conditions with a pwv of only 0.2~mm during the observations whereas that in each of the seven long baseline EBs is higher due to the larger pwv of 0.4-0.5~mm (see Table~\ref{tab:obs_details}). In addition, the velocity where the noise is largest is always at 0~km~s$^{-1}$ in topocentric units but it shifts in the LSRK frame as the conversion between the two frames depends on e.g., the time and date of the observations. Therefore, the channel rms in the combined image is elevated over a larger velocity range than in the individual EBs. 

The varying weather conditions across the EBs not only affect the rms in the spectra, but they also affect the bandwidth over which the data are averaged for the bandpass calibration as part of the automated ALMA data reduction pipeline. The EBs where this bandwidth was small ($\lesssim 2$~km~s$^{-1}$) show a narrow peak of a few km~s$^{-1}$ in the rms spectra due to the telluric whereas those binned over a larger velocity range up to 9.6~km~s$^{-1}$ show a much wider peak where the noise is elevated. All in all, the resulting noise spectrum is complex with frequency dependent noise depending on e.g., the time between different EBs and under which weather conditions the data were taken. 

\subsection{Channel maps, moment maps, and spectra} \label{app:maps_and_spectra}
\begin{figure*}
\centering
\includegraphics[width=\hsize]{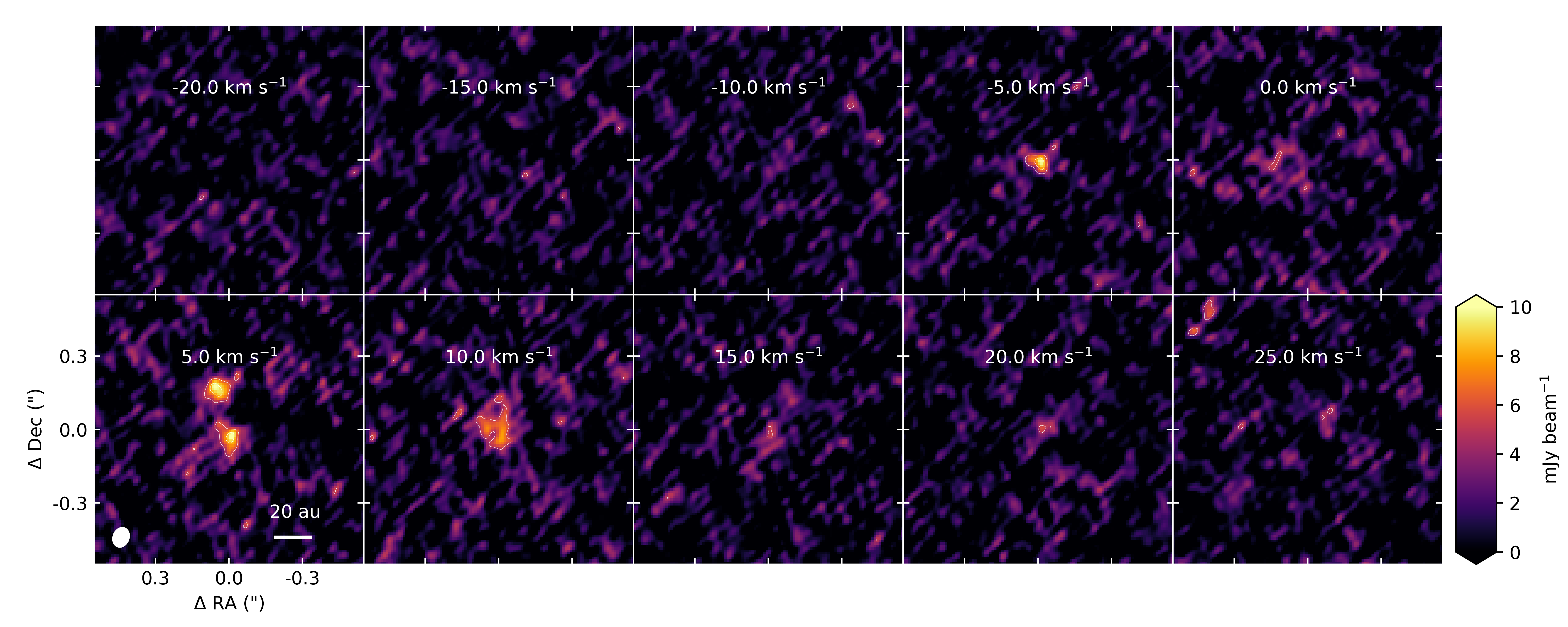}
  \caption{Channel maps of the \ce{H2O} line at 183~GHz imaged with a \texttt{robust} parameter of 0.0. The white contours indicate the $3\sigma$ and $5\sigma$ confidence levels where 1$\sigma$ corresponds to 1.5-1.7~mJy~beam$^{-1}$ depending on the velocity channel (see App.~\ref{app:telluric}). The beam and a 20~au scale bar are indicated in the bottom left panel. } 
     \label{fig:chans_H2O_r0.0}
\end{figure*}

\begin{figure*}
\centering
\includegraphics[width=\hsize]{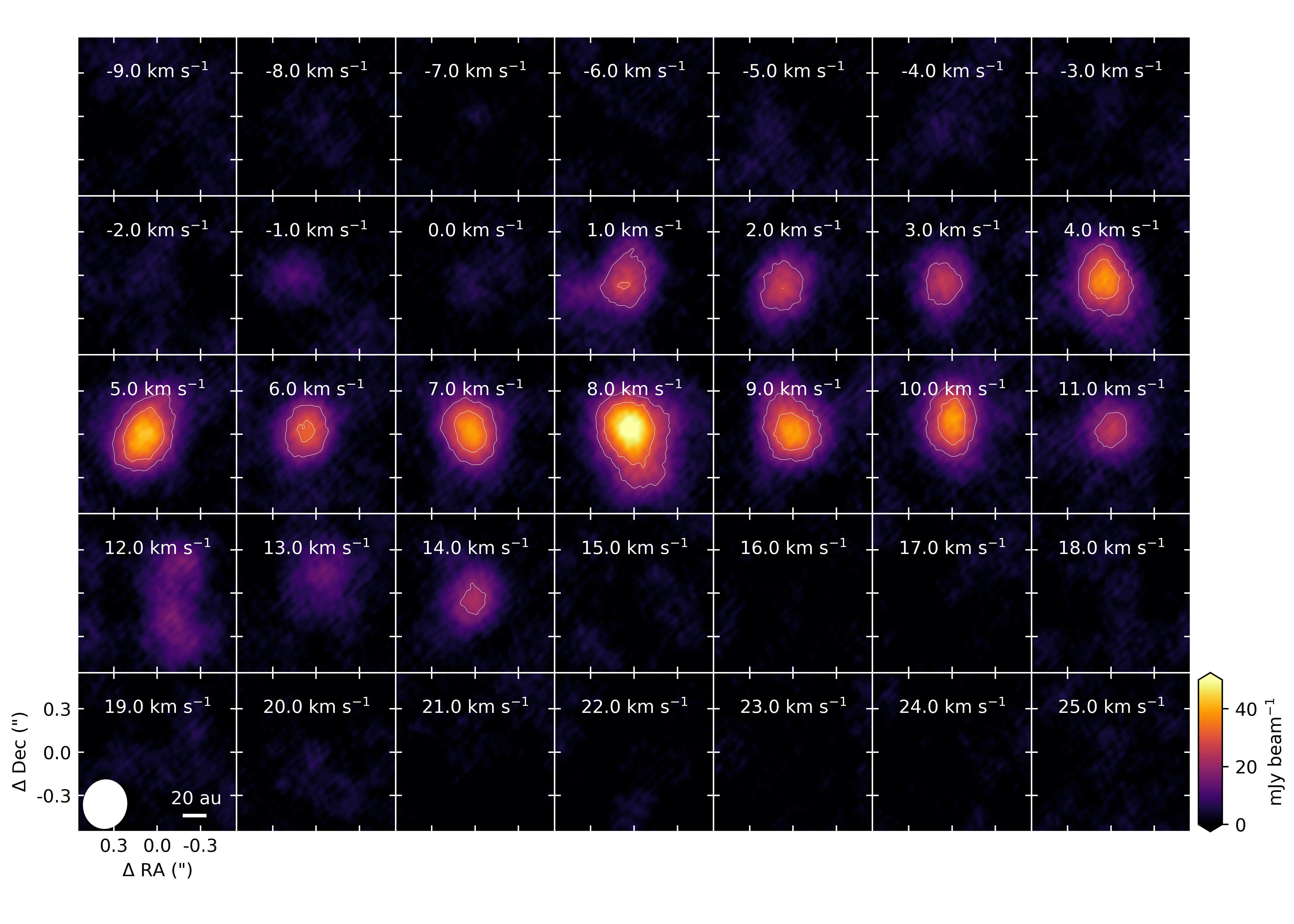}
  \caption{Channel maps of the \ce{H2O} line at 183~GHz imaged with a \texttt{robust} parameter of 2.0. The white contours indicate the $3\sigma$ and $5\sigma$ confidence intervals where 1$\sigma$ corresponds to 5-7~mJy~beam$^{-1}$ depending on the velocity channel (see App.~\ref{app:telluric}). The beam and a 20~au scale bar are indicated in the bottom left panel.}
     \label{fig:chans_H2O_r2.0}
\end{figure*}

\begin{figure*}
\centering
\includegraphics[width=\hsize]{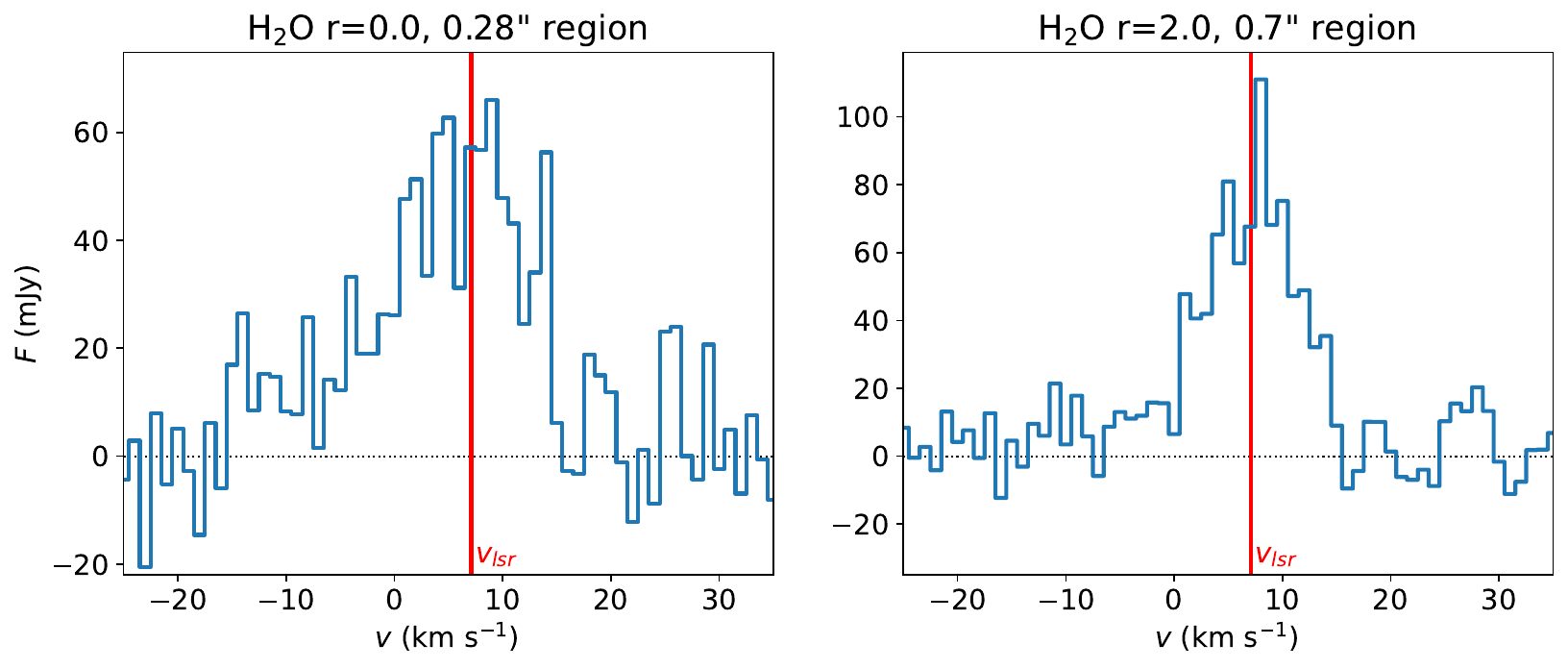}
  \caption{Spectra of the \ce{H2O} emission extracted from the same regions used to calculate the disk integrated flux without shifting and stacking the data. The left panel presents the spectrum of the line imaged with a \texttt{robust} parameter of 0.0 and the right panel the spectrum when imaging with a \texttt{robust} parameter of 2.0. The source velocity of the HL~Tau system is indicated by the red vertical line. }
     \label{fig:spec_H2O}
\end{figure*}

\begin{figure}
\centering
\includegraphics[width=\hsize]{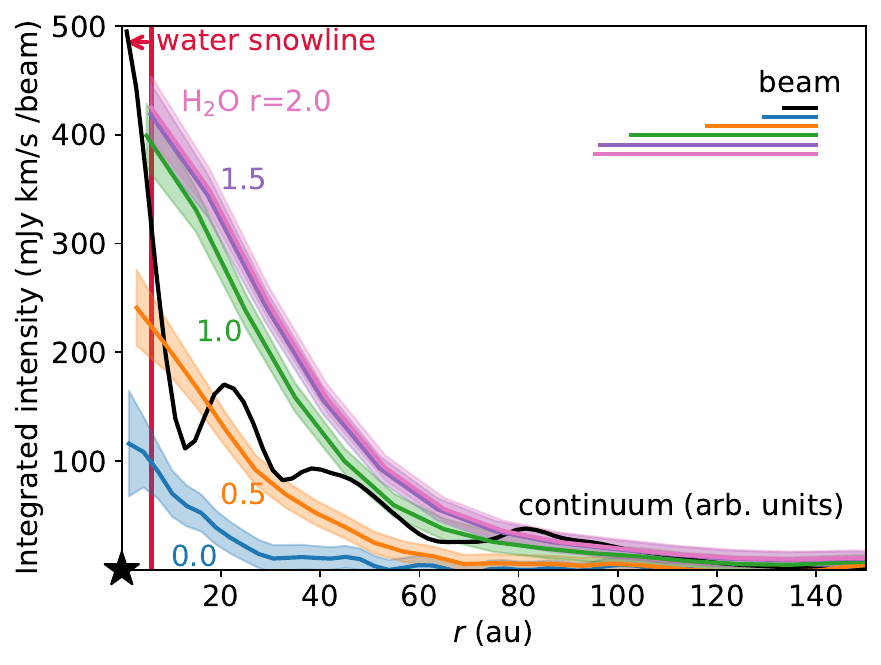}
  \caption{Azimuthally averaged radial profiles of the \ce{H2O} line at 183 GHz imaged with a \texttt{robust} parameter of 0 (blue), 0.5 (orange), 1 (green), 1.5 (purple), and 2.0 (pink) all corrected for the non-Gaussian beam shape, and the continuum emission (black). The continuum emission is presented in arbitrary units. The upper limit on the water snowline location is indicated with the vertical red line. The beams are indicated with the horizontal bars in the top right corner. }
     \label{fig:azi_avg_H2O_low_res}
\end{figure}

The channel maps of the \ce{H2O} 183~GHz line imaged at two different spatial resolutions are presented in Fig.~\ref{fig:chans_H2O_r0.0} and \ref{fig:chans_H2O_r2.0}. The $3\sigma$ and $5\sigma$ contours in both figures are computed using the rms noise in each channel separately following the App.~\ref{app:telluric}. The corresponding spectra extracted from a small $0\farcs28$ and larger $0\farcs7$ region are presented in Fig.~\ref{fig:spec_H2O}. 
The azimuthally averaged radial profile of the \ce{H2O} 183~GHz line imaged with a larger \texttt{robust} parameters of 0.0 up to 2.0 are presented in Fig.~\ref{fig:azi_avg_H2O_low_res}. All radial profiles are JvM-corrected to correct for the non-Gaussian shape of the beam affecting the diffuse, large scale emission. The \ce{H2O} emission shows a bright inner component and shelf of diffuse emission extending out to $\sim 75$~au that is only detected when imaged with a larger beam size than the \texttt{robust} = 0 image.

\begin{figure*}
\centering
\includegraphics[width=\hsize]{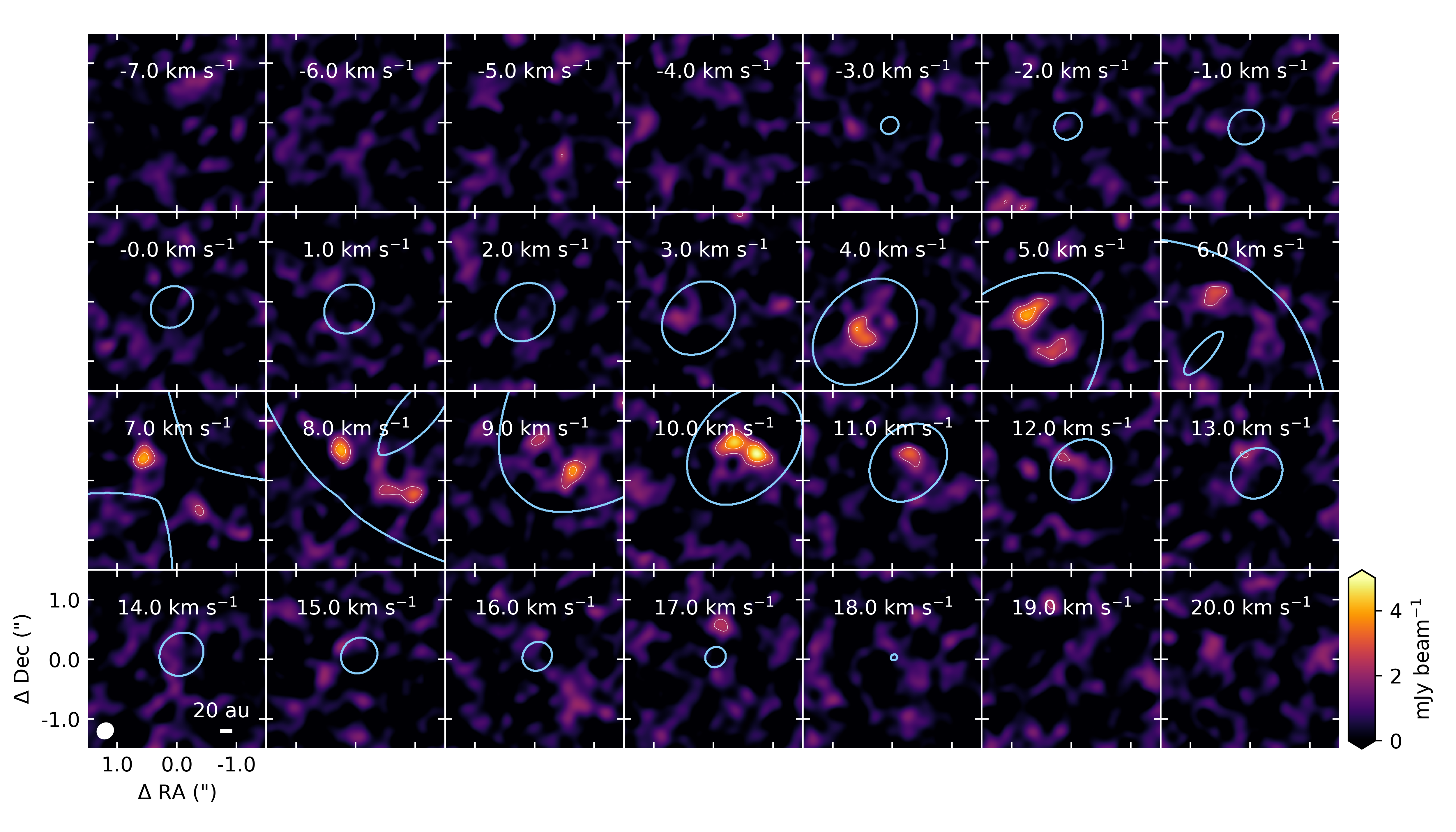}
  \caption{Channel maps of the \ce{H^13CO+} $J=2-1$ transition. The blue contours indicate the Keplerian mask used for the azimuthally averaged radial profile and the white contours indicate the $3\sigma$ and $5\sigma$ confidence levels where 1$\sigma$ corresponds to 0.7~mJy~beam$^{-1}$.} 
     \label{fig:chans_H13CO+}
\end{figure*}

\begin{figure*}
\centering
\includegraphics[width=\hsize]{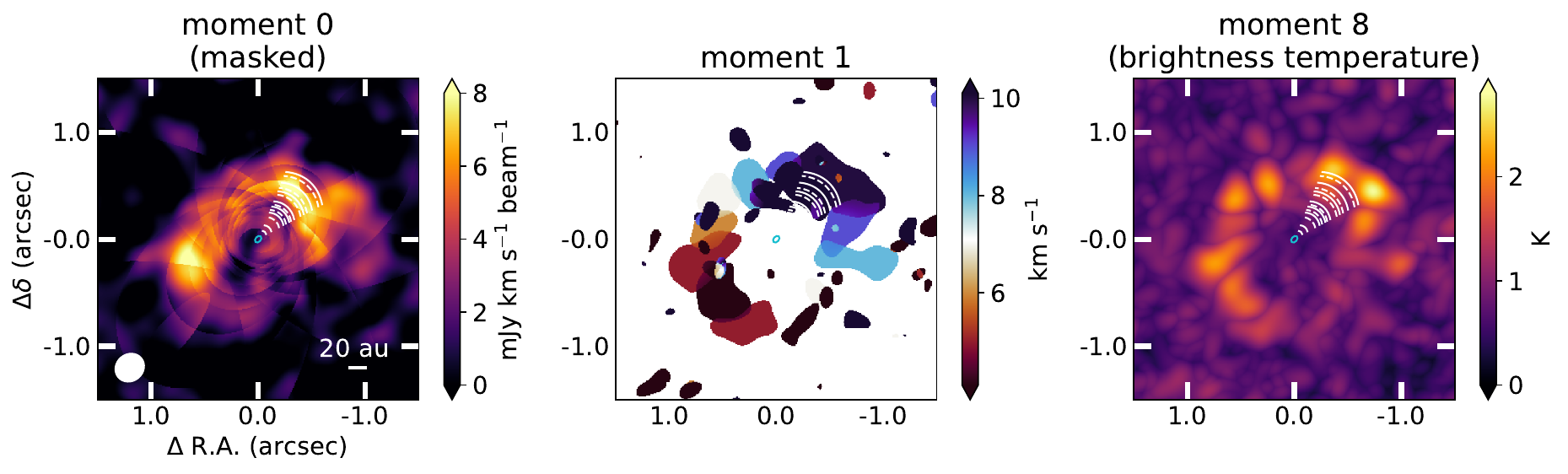}
  \caption{Moment 0, 1, and 8 maps of the \ce{H^13CO+} emission in the HL~Tau disk. A Keplerian mask was applied to the data before making the moment 0 map (left panel) but not before making the moment~1 and 8 maps (middle and right panel).}
     \label{fig:mom01_H13CO+}
\end{figure*}

\begin{figure*}
\centering
\includegraphics[width=\hsize]{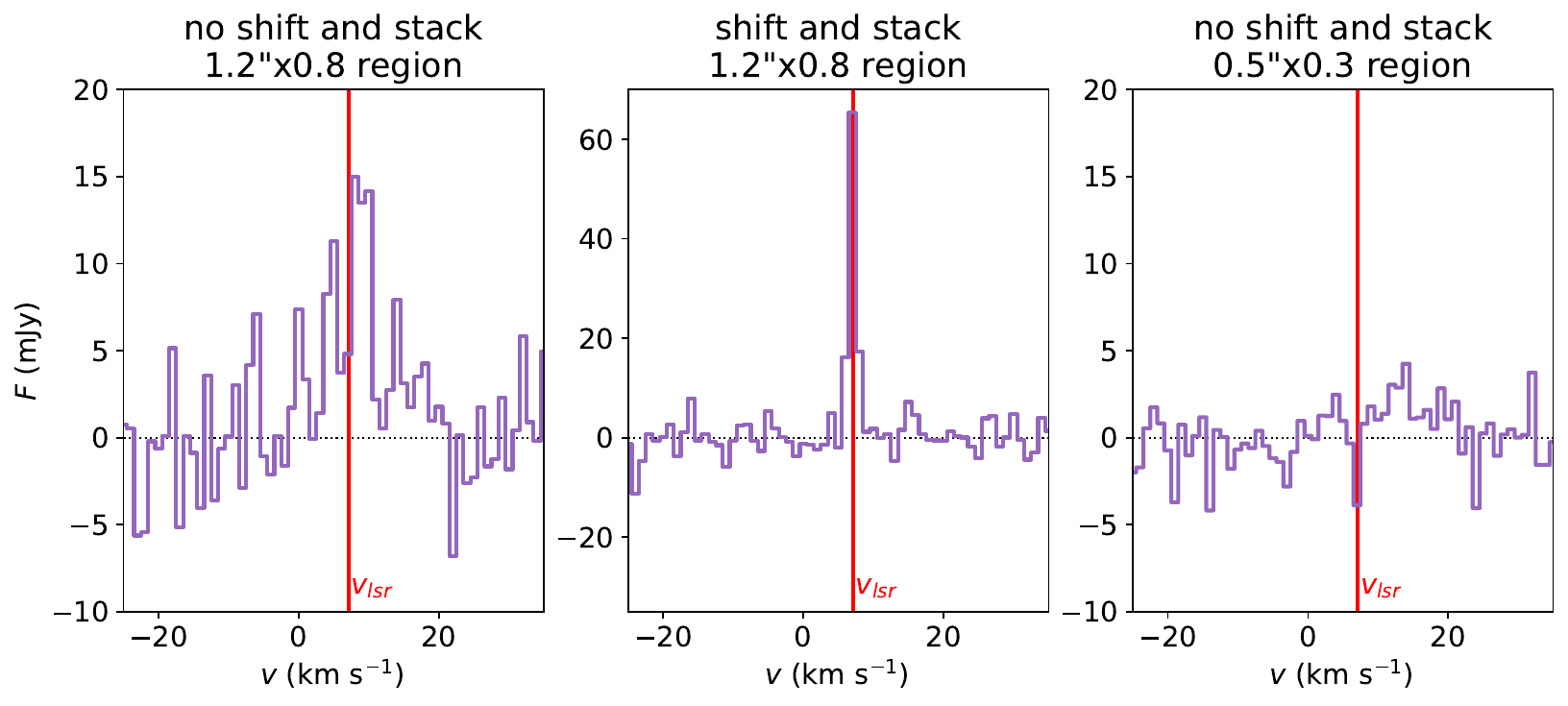}
  \caption{Spectrum of the \ce{H^13CO+} $J=2-1$ line extracted from a $1\farcs2 \times 0\farcs8$ elliptical region and an $0\farcs5\times 0\farcs3$ elliptical region positioned in the hole seen in the \ce{H^13CO+} moment maps. The left and right panel presents the spectrum without shifting and stacking and the middle panel shows the spectrum with shifting and stacking where each pixel is shifted by the projected Keplerian velocity at that location in the disk. The middle panel shows that the line is detected at the expected velocity.}
     \label{fig:spec_H13CO+}
\end{figure*}

The \ce{H^13CO+} $J=2-1$ channel maps are presented in Fig.~\ref{fig:chans_H13CO+}. The integrated intensity and the moment 1 and 8 maps of the \ce{H^13CO+} emission are presented in Fig.~\ref{fig:mom01_H13CO+}, where the moment~8 map is converted to brightness temperature using the Rayleigh-Jeans approximation. Even though the line is weak, a ring shaped profile is clearly visible when applying a Keplerian mask to the data before creating the integrated intensity map and in the moment~1 and 8 map without any masking applied. In addition, the moment~1 map shows a velocity gradient along the disk major axis, consistent with what is expected from a Keplerian rotating disk. Finally, the \ce{H^13CO+} spectrum extracted from a $1\farcs2\times 0\farcs8$ region is presented in the left panel of Fig.~\ref{fig:spec_H13CO+}. The middle panel represents the spatially integrated spectrum extracted after shifting each pixel by the local projected Keplerian velocity in the disk. This effectively removes the Doppler line broadening from the spectrum due to the Keplerian rotation, decreasing the linewidth and increasing the signal-to-noise ratio in the spectrum. The middle panel clearly shows that the \ce{H^13CO+} emission is detected at the expected velocity. The right panel shows the \ce{H^13CO+} spectrum extracted from a small elliptical region without shifting and stacking. No strong absorption is detected at the systemic velocity in the disk center, indicating that the hole seen in the \ce{H^13CO+} moment maps is not primarily due to absorption of emission in the envelope.

\begin{figure*}
\centering
\includegraphics[width=\hsize]{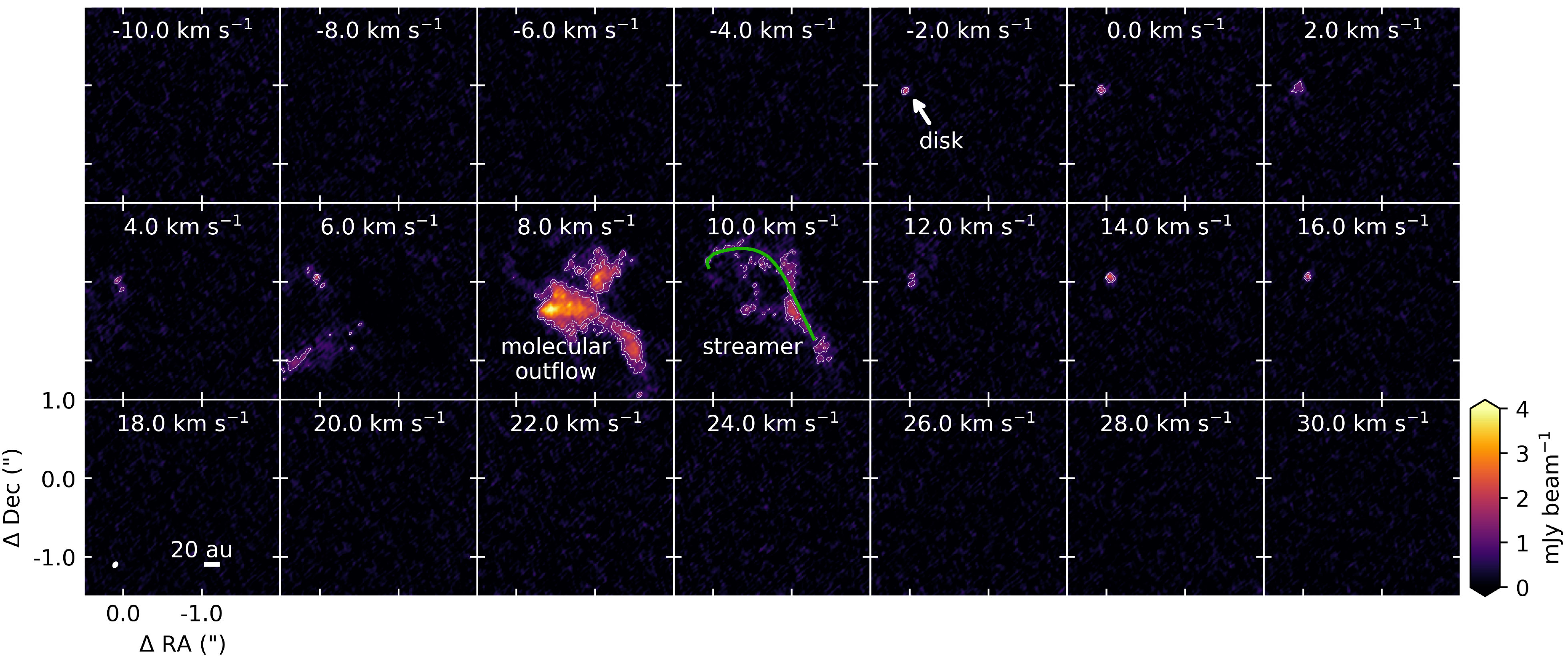}
  \caption{Channel maps of the SO $4_4-3_3$ line. The channels at 8 and 10~km~s$^{-1}$ trace primarily the the molecular outflow and primarily the streamer, respectively. The green line in the latter channel is the fit to the 1-arm spiral tracing the streamer in \ce{HCO+} \citep{Yen2019}. The white contours indicate the $3\sigma$ and $5\sigma$ confidence levels where 1$\sigma$ corresponds to 0.3~mJy~beam$^{-1}$. We note that the position of the disk is off center with respect to the center of the channel maps.}
     \label{fig:chans_SO}
\end{figure*}

\begin{figure*}
\centering
\includegraphics[width=\hsize]{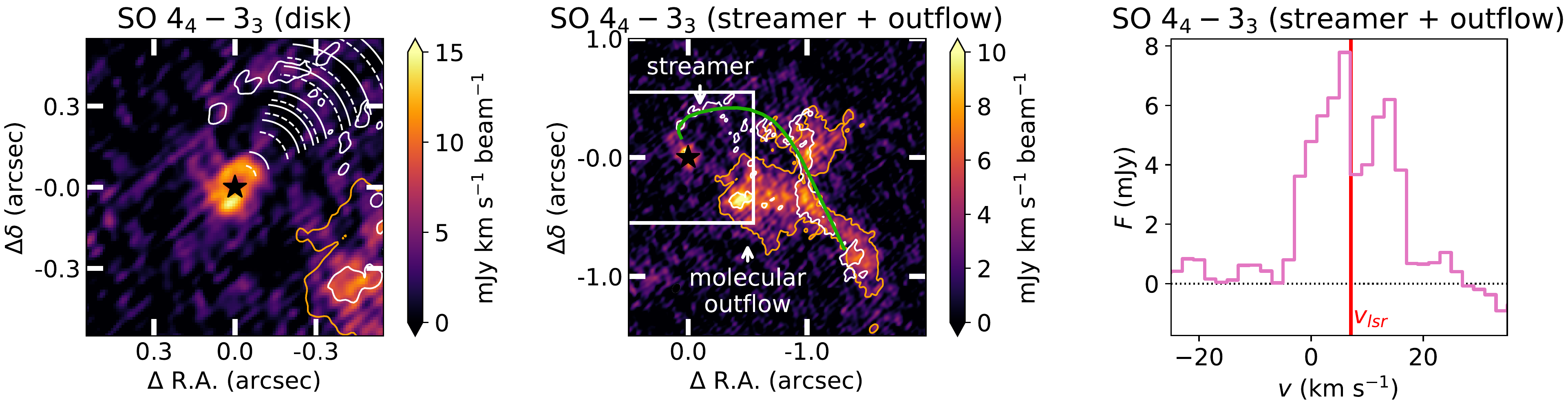}
  \caption{SO emission in the HL Tau system. Left and middle: integrated intensity maps of the SO emission. The left panel presents the map integrated from -5 to 19 km s$^{-1}$ zoomed in on the HL~Tau disk and the middle panel shows the emission integrated over a smaller velocity range from 5 to 13 km s$^{-1}$, highlighting the emission coming from primarily the streamer (white) and the molecular outflow (orange contour). The green line in the latter middle panel is the fit to the 1-arm spiral tracing the streamer in \ce{HCO+} \citep{Yen2019}. The right panel presents the SO spectrum. }
     \label{fig:mom0_SO}
\end{figure*}

The channel maps, integrated intensity maps, and the spectrum of the SO emission are presented in Fig.~\ref{fig:chans_SO} and \ref{fig:mom0_SO}. The left panel of the latter figure shows a zoom in of the disk whereas the middle panel shows the map integrated over a narrower velocity range of 5 to 13 km~s$^{-1}$ instead of -5 to 19 km~s$^{-1}$ to highlight the SO streamer and molecular outflow. Additionally, the SO spectrum is presented in the right panel. 

\section{Thermochemical model} \label{app:DALI}

To investigate the fraction of the water reservoir accessible to observations at various wavelengths, we constructed a simple thermochemical model reproducing the rare CO isotopologue lines seen with ALMA and the 0.95~mm continuum emission within a factor of 2 in most disk regions. We focus on the \ce{C^18O}, \ce{^13C^18O}, and \ce{^13C^17O} $J=3-2$ lines seen with ALMA to trace the HL~Tau disk and minimize the effect of the surrounding envelope and molecular outflow seen in e.g., CO \citep{Furlan2008, Lumbreras2014, Bacciotti2025}. 

The disk is modelled using the thermo-chemical modelling code DALI \citep{Bruderer2009, Bruderer2012, Bruderer2013}. We use the standard model for a full disk that is described by parameters for a viscously evolving disk between 0.23 and 500~au \citep{Lynden-Bell1974, Hartmann1998}:
\begin{align}
    \Sigma_{\rm gas} = \Sigma_c \left (\frac{r}{r_c} \right )^{-\gamma} \exp{\left [-\left (\frac{r}{r_c} \right )^{2-\gamma}\right ]}
\end{align}
with $\Sigma_c$ setting the surface density at the characteristic radius $r_c$ and $\gamma = 1$ the power law index of the profile. We set the total disk mass to 0.2~M$_{\odot}$ based on the \ce{^13C^17O} detection in this disk \citep{Booth2020} and the characteristic radius to 30~au to roughly match the morphology of the CO isotopologue emission. The disk scale height $h_c$ is set to 0.1 \citep{Pinte2016}, whereas the grains are settled to the disk midplane with a scale height of $0.05\times h_c$ as the dust disk is generally flat \citep{GuerraAlvarado2024, Yang2025}. The disk flaring is parametrized as $h=h_c(r/r_c)^{\psi}$, with a flaring index of 0.15.

The stellar spectrum is modelled as a 4000~K black body \citep{Liu2017}, where the mass accretion rate of $8.7\times 10^{-8}$~M$_{\odot}$~yr$^{-1}$ is modelled as an additional 10000~K black body spectrum \citep{Beck2010}. The X-ray luminosity of the star is $3.36\times 10^{30}$~erg~s$^{-1}$ \citep{Skinner2020} and a cosmic ray ionization rate of $10^{-17}$~s$^{-1}$ is assumed. The stellar mass is set to 2.1~M$_{\odot}$ \citep{Yen2019}. 

Initially the model is run time independently with a small chemical network consisting of 109 species and 1463 reactions. This network is sufficient to obtain the temperature structure of the disk and includes the \ce{H2O} self-shielding \citep{Bosman2022}. Then the model is either run with the same network time dependently for 1~Myr, the approximate age of the disk \citep{Briceno2002} to obtain the water abundance or it is run time dependently with the CO isotopologue network described in \citet{Miotello2016} to model the emission of the CO isotopologues. 

The CO isotopologue network is suited to model the emission of rare CO isotopologues as isotope selective effects of single CO isoptopologues are taken into account as well as the self-shielding of the individual isotopologues. Mutual self-shielding is not included. This network uses a \ce{^12C}/\ce{^13C} ratio of 70, a \ce{^16O}/\ce{^18O} ratio of 560, and a \ce{^18O}/\ce{^17O} ratio of 3.6 \citep{Wilson1999, Milam2005}.

\begin{figure*}
\centering
\includegraphics[width=\hsize]{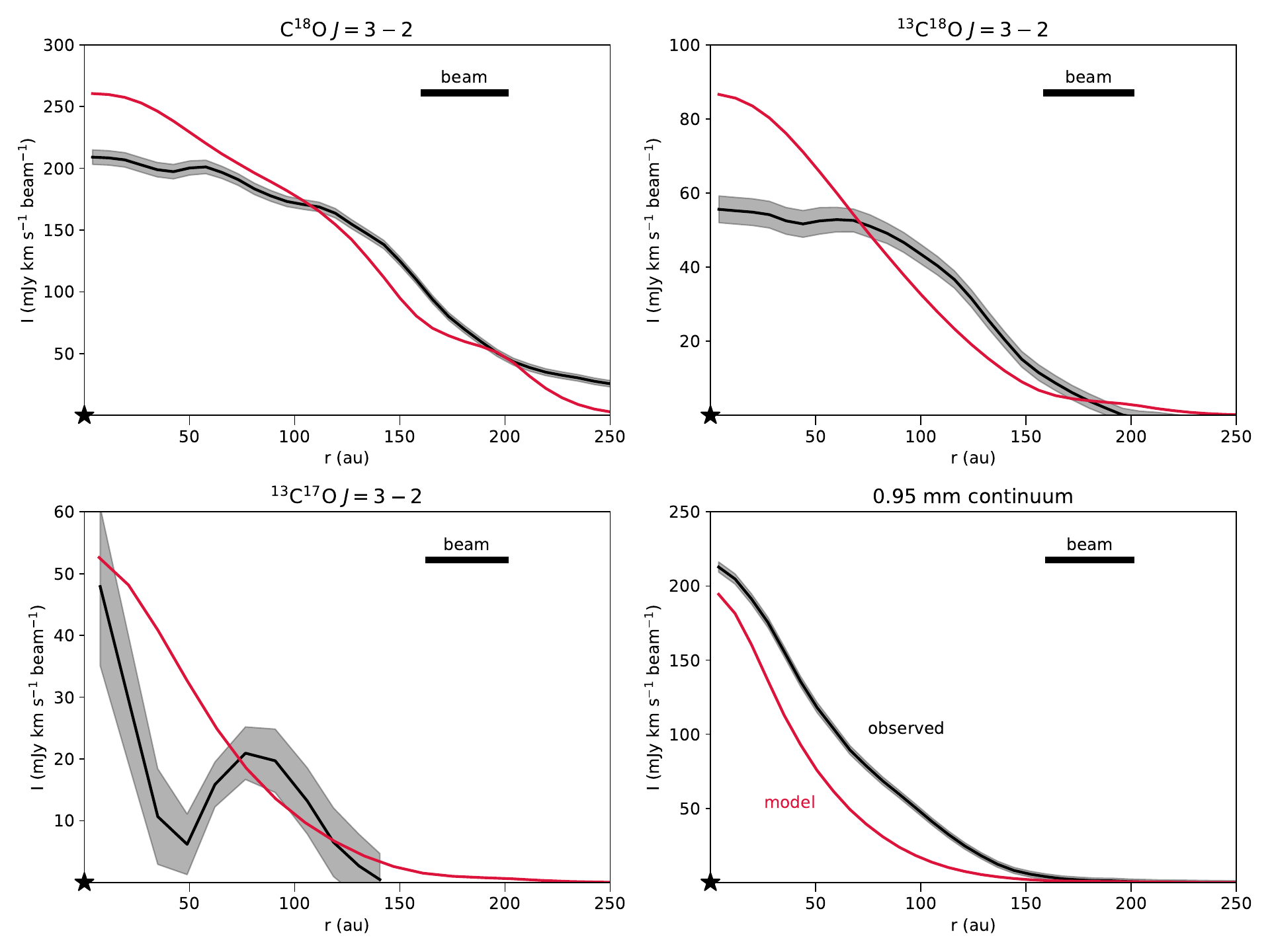}
  \caption{Azimuthally averaged radial profiles of the \ce{C^18O}, \ce{^13C^18O}, and \ce{^13C^17O} $J=3-2$ transition in the HL~Tau disk together with the 0.95~mm continuum. The data are shown in black and are taken from \citep{Booth2020} and the product data from ALMA project 2021.1.01310.S (PI: K. Zhang). The model predictions are shown in red. }
     \label{fig:dali_CO_dust}
\end{figure*}

Figure~\ref{fig:dali_CO_dust} presents the resulting azimuthally averaged radial profiles of the \ce{C^18O}, \ce{^13C^18O}, and \ce{^13C^17O} $J=3-2$ emission predicted by the CO isotopologue network and the 0.95~mm continuum emission. These profiles are compared to the observed profiles from \citet{Booth2020} and the product data from ALMA program 2021.1.01310.S (PI: K. Zhang) that will be presented in detail in \citet{TorresVillanuevainprep}. The modelled radial profiles reproduce the data within a factor of $<2$ for most disk radii at the resolution of the data.

\end{document}